\documentclass[prd, aps, twocolumn, showpacs, superscriptaddress,floatfix]{revtex4}
\usepackage{graphicx}
\usepackage{bm}
\newcommand{\mres}{m_{\rm res}}
\newcommand\riken{RIKEN-BNL Research Center, Brookhaven National Laboratory, Upton, NY 11973, USA}
\newcommand\bnlaf{Brookhaven National Laboratory, Upton, NY 11973, USA}
\newcommand\edinb{SUPA, School of Physics, The University of Edinburgh, Edinburgh EH9 3JZ, UK}
\newcommand\epcca{EPCC, School of Physics, The University of Edinburgh, Edinburgh EH9 3JZ, UK}
\newcommand\cuaff{Physics Department, Columbia University, New York, NY 10027, USA}
\newcommand\glasg{SUPA, Department of Physics \& Astronomy, University of Glasgow, Glasgow G12 8QQ, UK}
\newcommand\swans{Department of Physics, University of Wales Swansea,
  Swansea SA2 8PP, UK}
\newcommand\kanaz{Institute for Theoretical Physics,  Kanazawa University, Kakuma, Kanazawa, 920-1192, Japan}
\newcommand\tokyo{Department of Physics, University of Tokyo, Hongo 7-3-1, Bunkyo-ku, Tokyo 113, Japan}
\newcommand\uconn{Physics Department, University of Connecticut, Storrs, 
Connecticut 06269-3046, USA}

\begin{document}
\bibliographystyle{apsrev}

\title{2+1 flavor domain wall QCD on a (2 fm)$^3$ lattice: \\light meson spectroscopy with $L_s$ = 16}
\author{C.~Allton}      \affiliation{\swans}
\author{D.J.~Antonio}   \affiliation{\edinb}
\author{T.~Blum}        \affiliation{\riken}\affiliation{\uconn}
\author{K.C.~Bowler}    \affiliation{\edinb}
\author{P.A.~Boyle}     \affiliation{\edinb}
\author{N.H.~Christ}    \affiliation{\cuaff}
\author{S.D.~Cohen}     \affiliation{\cuaff}
\author{M.A.~Clark}     \affiliation{Center for Computational Science, 3 Cummington Street, Boston University, MA 02215, USA}
\author{C.~Dawson}      \affiliation{\riken}
\author{A.~Hart}	\affiliation{\edinb}
\author{K.~Hashimoto}   \affiliation{\kanaz}\affiliation{Radiation Laboratory, RIKEN, Wako, Saitama 351-0198, Japan}
\author{T.~Izubuchi}    \affiliation{\riken}
\affiliation{\kanaz}
\author{A.~J\"uttner}    \affiliation{School of Physics and Astronomy, University of Southampton,  Southampton SO17 1BJ, UK}
\author{C.~Jung}        \affiliation{\bnlaf}
\author{A.D.~Kennedy}   \affiliation{\edinb}
\author{R.D.~Kenway}    \affiliation{\edinb}
\author{M.~Li}         \affiliation{\cuaff}
\author{S.~Li}         \affiliation{\cuaff}
\author{M.F.~Lin}         \affiliation{\cuaff}
\author{R.D.~Mawhinney} \affiliation{\cuaff}
\author{C.M.~Maynard}   \affiliation{\epcca}
\author{J.~Noaki\footnote{Present address: Institute of Particle and Nuclear Studies, KEK, Ibaraki 305-0801, Japan}}
  \affiliation{School of Physics and Astronomy, University of Southampton,  Southampton SO17 1BJ, UK}
\author{S.~Ohta}        \affiliation{\riken}
\affiliation{Institute of Particle and Nuclear Studies, KEK, Ibaraki 305-0801, Japan}
\affiliation{The Graduate University for Advanced Studies (Sokendai), Tsukuba, Ibaraki 305-0801, Japan}
\author{B.J.~Pendleton}   \affiliation{\edinb}
\author{S.~Sasaki} \affiliation{\riken}\affiliation{\tokyo}
\author{E.E.~Scholz} \affiliation{\bnlaf}
\author{A.~Soni}        \affiliation{\bnlaf}
\author{R.J.~Tweedie}   \affiliation{\edinb}
\author{A.~Yamaguchi}   \affiliation{\glasg}
\author{T.~Yamazaki}   \affiliation{\uconn}
\collaboration{RBC and UKQCD Collaborations}
%
%
\noaffiliation{RBC and UKQCD Collaborations}


\date{\today}

\begin{abstract}
We present results for light meson masses and pseudoscalar decay
constants from the first of a series of lattice calculations with 2+1 dynamical
flavors of domain wall fermions and the Iwasaki gauge action. The work
reported here was done at a fixed lattice spacing of about 0.12 fm on
a $16^3\times32$ lattice, which amounts to a spatial volume of (2
fm)$^3$ in physical units. The number of sites in the fifth dimension is 
16, which gives  $m_{\rm res} =
0.00308(4)$ in these simulations. Three values of input light sea
quark masses, $m_l^{\rm sea} \approx 0.85\, m_s, 0.59 \,m_s$ and $0.33
\,m_s$ were used to allow for extrapolations to the physical light
quark limit, whilst the heavier sea quark mass was fixed to
approximately the physical strange quark mass $m_s$. The exact
rational hybrid Monte Carlo algorithm was used to evaluate the
fractional powers of the fermion determinants in the ensemble
generation.  We have found that $f_\pi = 127(4)$ MeV, $f_K = 157(5)$
MeV and $f_K/f_\pi = 1.24(2)$, where the errors are statistical only,
which are in good agreement with the experimental values. 
\end{abstract}

\maketitle


\section{Introduction}
\label{sec:Introduction}

The successful determination of the light hadron spectrum serves
as an important test of lattice QCD. With the emergence of powerful
computers and new algorithms, lattice QCD has entered an era of
dynamical simulations where two light quarks and one strange quark
are included in the vacuum polarization effects.  (These simulations
are referred to as 2+1 flavor simulations.)  Since current
simulations use light quarks that are heavier than the physical
values, extrapolations to the light quark region are still required,
and having control of these extrapolations is vital for a
comparison with experimental results.  Chiral perturbation theory
provides a framework for such extrapolations, although the range of quark masses
for which a given order of chiral perturbation theory is accurate is
still under investigation.

Symmetry plays an important role in hadron physics. The domain wall
fermion (DWF) formulation~\cite{Kaplan:1992bt,Shamir:1993zy,Furman:1994ky}
respects flavor symmetry and has approximate chiral symmetry, with
the introduction of an auxiliary fifth dimension.  When the extent
of this fifth  dimension, denoted as $L_s$, goes to infinity, chiral
symmetry is fully recovered.  In addition, unlike Wilson and staggered
fermions, domain wall fermions (and the closely connected overlap
fermions) allow chiral symmetry to be recovered at finite lattice
spacing.  While certainly an important improvement for the light hadron
spectrum calculation presented here, this good chiral symmetry  
is vital for accurate operator renormalization and control of 
operator mixing in a wide
variety of important hadronic matrix elements.  As evidenced by
existing work in quenched QCD, the chiral properties of domain wall
fermions make calculations possible that cannot currently be done
with other formulations \cite{Blum:2001xb, Noaki:2001un}.  To move
beyond the quenched approximation, 2+1 flavor ensembles are needed.
The production of these ensembles and the determination of their basic
properties are important steps towards the goal of measuring a wide
variety of physically interesting quantities using this formulation
and are the focus of this paper.

In practice, $L_s$ is taken to be on the order of 10-20 sites so that some
residual chiral symmetry breaking remains.  However, this residual chiral
symmetry breaking is highly suppressed.  Compared to Wilson fermions
where the residual chiral symmetry breaking is an ${\cal O}(1)$ effect, 
for DWF it is of ${\cal O}(10^{-3})$ or smaller, depending on the size of $L_s$.  
To leading order in the lattice spacing $a$, the only effect of this residual chiral 
symmetry breaking is a simple mass renormalization (denoted by $\mres$)
added to the input quark mass.  Additional chiral symmetry breaking effects
enter at higher order in $a$, but they are suppressed by the additional
powers of $a$ and also vanish as $L_s$ becomes large.  This situation
simplifies the chiral extrapolation of physical observables, and allows 
us to study those aspects of hadron phenomenology where chiral symmetry 
is important with reduced systematic uncertainties.

Domain wall fermions are both on- and off-shell improved, since for
any quantity a traversal of the fifth dimension is required to mix
chiralities.  Such a traversal is suppressed for large $L_s$, a suppression 
which gives rise to the small value for $m_{\rm res}$ and which depends 
on $L_s$ as $\exp(-\alpha L_s)$ if the generic, weak coupling behavior of 
residual chiral symmetry breaking effects is used \cite{Golterman:2004mf}.
Thus, the resulting theory is accurate up to terms of ${\cal O}(
a \exp(-\alpha L_s))$ and ${\cal O}(a^2)$.

For amplitudes which violate chiral symmetry by two units, a
suppression of order $m_{\rm res}^2$ or $\exp(-2\alpha L_s)$ is
expected.  This estimate may fail in the case of composite operators 
with many fields at the same space-time point\cite{Golterman:2004mf}, 
where one must give special consideration to chiral symmetry breaking 
arising from rare, localized modes that are undamped in the fifth dimension.  
However, the detailed flavor structure of the operator must be considered 
which, in important cases, prevents these localized modes from contributing
\cite{Christ:2005xh}.  The result is that chirally disallowed operator 
mixings are generally under good control with DWF at finite $L_s$.

For power divergent quantities, one finds that if an operator receives 
an $m_f/a^n$ contribution, where $m_f$ is the quark mass, then for finite 
$L_s$ it will generally receive an ${\cal O}(\mres)/a^n$ contribution 
\cite{Blum:2000kn}.  Because the residual chiral symmetry breaking is 
entering a divergent amplitude, it will not be determined by the chiral 
symmetry breaking at low energies described by $\mres$.  However, it will
be of this order, as has been demonstrated in \cite{Blum:2001xb}.   Thus,
for such quantities the simple replacement $m_f/a^n \rightarrow (m_f+m_{\rm res})/a^n$
provides a sensible estimate of but not an accurate value for the effects
of residual symmetry breaking on the quantity in question.  For example,
such a term enters the chiral condensate $\langle \bar{q}q \rangle$.
This implies that to compute $\langle \bar{q}q \rangle$ with domain wall
fermions an extrapolation to large $L_s$ must be made or that this physical
quantity should be determined from the density of Dirac eigenvalues at 
zero, so that ultraviolet chiral symmetry breaking effects do not enter.
In contrast, a similar, uncertain ${\cal O}(m_{\rm res}/a^2)$ term which 
appears in the matrix elements of the weak interaction operator $O_6$, 
does not effect the quantity of interest which is proportional to $m_f$ 
\cite{Blum:2001xb}.  For a more complete discussion of the effects of 
residual chiral symmetry breaking for domain wall fermion see, for example, 
Refs~\cite{Furman:1994ky,Blum:2000kn,Blum:2001sr,Blum:2001xb}.

These benefits of controlled chiral symmetry breaking must be weighed 
against the computing cost for domain wall fermions, which are naively 
${\cal O}(L_s)$ more expensive than conventional fermion 
formulations~\cite{Clark:2006wq}.  In balance, we believe that the 
benefits of domain wall fermions outweigh their extra numerical costs, 
because of the much larger range of observables that are accessible to 
this formulation.

Studies which established a set of parameter values suitable for
2+1 flavor simulations with domain wall fermions were reported
in~\cite{Antonio:2006px}.  In the present paper we describe the first of a series of 2+1
flavor DWF simulations which use the parameters as determined in
\cite{Antonio:2006px}, and explore in detail the 
systematic effects arising in such a full dynamical simulation with
domain wall fermions.  Specifically, we use the Iwasaki gauge
action with gauge coupling $\beta = 2.13$ on a lattice of $16^3
\times 32$, which gives a lattice spacing of about 0.12 fm and a
spatial lattice extent of about 2 fm.  This volume should be large 
enough to give accurate results for the physics of the pseudoscalar 
mesons simulated here: those with a mass ratio to vector meson
masses in the range $0.5 \le m_P/m_V \le 1.0$.
Larger volume simulations at the same coupling using lighter quark
masses are underway, as are simulations at weaker couplings, also
with lighter quark masses.  The volumes used in the work presented
here do not allow lighter dynamical quark masses to be used, so
the full benefits of DWF will not be visible.  However, the current
simulations allow us to demonstrate that more costly simulations at
larger volumes and lighter quark masses are feasible. Since
baryonic observables may suffer from non-negligible finite volume
effects in the current volume, we will only focus on light meson
physics in this paper.  In particular, we will present results for
light pseudoscalar meson masses and decay constants, and calculate
quantities of direct phenomenological interest, such as the light
quark masses and $f_K/f_\pi$.

One of the many complexities of 2+1 flavor simulations is the presence
of the fractional powers of the fermion determinants in the path
integral, which the conventional hybrid Monte Carlo algorithm is not
able to evaluate. We employed the rational hybrid Monte Carlo (RHMC) 
algorithm by Clark and
Kennedy~\cite{Clark:2003na,Clark:2004cp,Clark:2005sq}, which is not
only free of finite step size errors, but also has proved to be quite
efficient after recent improvements and
tunings~\cite{Mawhinney:lat06}. The gauge configurations used
in this work were mainly generated with one variant of the RHMC
algorithm. We describe the details of this algorithm in Section~\ref{sec:simulation_details}.
The algorithm was further improved after the major part of this work
had been done. It is thus to our advantage to study how the computing
cost and ensemble properties change with the new improvements. We
compare these two variants of the RHMC algorithm in Section~\ref{sec:quotient_RHMC}. 

This paper is organized as follows. In
Section~\ref{sec:simulation_details} we present the details of
generation of the gauge configurations using the RHMC
algorithm.  We discuss the thermalization and autocorrelations
of  the ensembles in Section~\ref{sec:ensemble_properties}.
Section~\ref{sec:observables} gives the results of the light meson masses
and  decay constants, including the residual chiral symmetry breaking
in these simulations. The results for  physical observables in the
chiral limit are presented in Section~\ref{sec:chiral_limit}.
Section~\ref{sec:quotient_RHMC} discuses an extension of one of the
ensembles using an improved RHMC algorithm. Our conclusions are given
in Section~\ref{sec:conclusions}.

\section{Generation of Gauge Configurations}
\label{sec:simulation_details}

In this study we have used the Iwasaki renormalization-group improved
gauge action and $2+1$ flavors of dynamical domain wall fermions;
details of the action and our notation may be found in~\cite{Antonio:2006px}. Our
ensembles were generated using two new variants of the
RHMC algorithm as described below.

For the case of two degenerate light quarks of mass, $m_l$, and a
strange quark of mass, $m_s$, integrating out the fermion fields in
the path integral leads to
\begin{widetext}
\begin{equation}
\frac{
 \det \left[ D_{\rm DWF}^\dagger(M_5, m_l) D_{\rm DWF}(M_5, m_l) \right]
 }{
 \det \left[ D_{\rm DWF}^\dagger(M_5, 1) D_{\rm DWF}(M_5, 1) \right]
 }
\;\;
\frac{
 \det^{1/2} \left[ D_{\rm DWF}^\dagger(M_5, m_s) D_{\rm DWF}(M_5, m_s)
   \right] }{
 \det^{1/2} \left[ D_{\rm DWF}^\dagger(M_5, 1) D_{\rm DWF}(M_5, 1)
    \right]
 }
 \label{eq:nf3-det}
\end{equation}
\end{widetext}
where $D_{\rm DWF}(M_5, m)$ is the domain wall Dirac operator, $M_5$
is the height of the domain wall, and $m$ is the fermion mass. The
determinants in the denominator arise from the Pauli-Villars fields
introduced to cancel the bulk infinity which arises as the size of
the fifth dimension $L_s\to\infty$.

We may use the identity $\det(M) = [\det^{1/n}(M)]^n$ to simulate
various different decompositions of the same fermionic determinants
using RHMC.  If we adopt the shorthand notation ${\cal D}(m_i) =
D_{\rm DWF}^\dagger(M_5, m_i) D_{\rm DWF}(M_5, m_i)$, we may write the
ratio of determinants in Eq.~(\ref{eq:nf3-det}) as

\begin{equation}
\frac{
  \det^{1/2}[{\cal D}(m_l)] \; \det^{1/2}[{\cal D}(m_l)] \;
  \det^{1/2}[{\cal D}(m_s)]
}{
  \det[{\cal D}(1)] \; \det^{1/2}[{\cal D}(1)]
}.
\label{eq:DeterminantDecomp}
\end{equation}

In our previous paper~\cite{Antonio:2006px}, each determinant in
Eq.~(\ref{eq:DeterminantDecomp}) was associated with a separate
pseudo-fermion field, and a standard RHMC
algorithm~\cite{Kennedy:1998cu,Clark:2006fx} was employed.

The pseudo-fermion action for molecular dynamics evolution of a single
quark flavor of mass $m$ takes the form $S_{\rm pf} = \phi^\dagger
{\cal D}^{-1/2}(m)\phi$. We may improve the efficiency of RHMC by
taking advantage of the properties of the expansion coefficients in
the partial-fraction formulation of the $p^{\rm th}$-order rational
approximation, \(x^{-1/2} \simeq \left\{\alpha_0 + \sum_{k=1}^p
\frac{\alpha_k}{x+\beta_k}\right\}\). The ``shifts'', $\beta_k$, which
are all real and positive, are ordered according to increasing
magnitude, and the $\alpha_k$ are all real and positive.  The fermion
force is
\begin{widetext}
\begin{equation}
-S'_{\rm pf} = \sum_{k=1}^p \alpha_k \, \phi^\dagger \left({\cal
  D}(m)+\beta_k\right)^{-1} {\cal D}'(m) \left({\cal
  D}(m)+\beta_k\right)^{-1} \phi \,,
\label{eq:fermion-force}
\end{equation}
\end{widetext}
where the prime denotes differentiation with respect to the gauge
field.  Somewhat remarkably, the inverses with the smallest shifts,
$\beta_k$, which are the most expensive to evaluate, are associated
with the smallest residues $\alpha_k$, and thus lead to
relatively-small contributions to the fermion
force~\cite{Clark:2006fx,Clark:2005sq}. It is then natural to split
the sum in Eq.~(\ref{eq:fermion-force}) into two parts: ``light
poles'' with $1\le k \le r$, and ``heavy poles'' with $r < k \le p$,
and evaluate their contributions to the fermion force on different
timescales using a Sexton-Weingarten multi-timescale
integrator~\cite{Sexton:1992nu}.  The light-pole contributions are
evaluated only on the largest timescale, the heavy poles on an
intermediate timescale, and the gauge fields on the finest timescale.
\begin{table}[ht]
\caption{Parameter values for $2+1$ flavor RHMC runs on $16^3\times32$
lattices with $L_s=16$, $M_5=1.8$, $\beta=2.13$, $m_s^{
\rm sea}=0.04$. $N_{\rm  MD \; units}$ is
the number of  molecular dynamics time units obtained from an ordered
start, except for the $m_l^{\rm sea} = 0.03$ ensemble generated using
RHMC II, which started from the last configuration generated using
RHMC I.  $N_{\rm step1}$ is the number of coarse integration steps of
size $\delta\tau$ per MD time unit. The intermediate integration scale is
$\delta\tau/N_{\rm step2}$ and the finest scale is
$\delta\tau/(N_{\rm step2}\, N_{\rm step3})$. The RHMC acceptance rate
is given in the second last column. The final column denotes the variant of
the RHMC algorithm used. 
\label{tab:run-details}}
\begin{tabular*}{\columnwidth}{@{\extracolsep{\fill}}cccccccc}
\hline
\hline
$m_l^{\rm sea}$ & $N_{\rm MD \; units}$ & $\delta\tau$ &
$N_{\rm step1}$ & $N_{\rm step2}$ & $N_{\rm step3}$ & Acc & Alg\\
\hline
0.01 & 4015 & 0.13 & 8 & 2 & 8 & 70\% & RHMC I\\
0.02 & 4045 & 0.14 & 7 & 2 & 8 & 56\% & RHMC I\\
0.03 & 4020 & 0.14 & 7 & 2 & 8 & 57\% & RHMC I\\
\hline
\hline
0.03 & 3580 & 0.25 & 4 & 1 & 6 & 80\% & RHMC II\\
\hline
\hline
\end{tabular*}
\end{table}

In order to maximize the acceptance rate per unit computational cost,
we optimized the order of the rational approximations and the points,
$r$, at which the partial-fraction sums are split, whilst also varying
the three Sexton-Weingarten timescales. The optimal rational
approximations and splits were $(p,r)=(10,4)$ for the light quarks,
$(p,r)=(10,3)$ for the strange quark, and $(p,r)=(6,1)$ for the
Pauli-Villars field associated with the strange quark.

In the RHMC accept/reject step, we used rational approximations of
order $16$ and $9$, respectively, for the quark and Pauli-Villars
fields.

For all ensembles, we used the second-order minimum-norm (2MN)
integrator~\cite{Sexton:1992nu,Omelyan:2003,Takaishi:2005tz,Urbach:2005ji} with
approximately unit-length trajectories. We tuned the free parameter in
2MN to $\lambda=0.22$ to minimize the RMS change in the Hamiltonian, $\langle (\Delta H)^2 \rangle ^{1/2}$, along a set of RHMC trajectories.

This completes the description of the {RHMC~I} algorithm used to
generate the ensembles in the main part of this work. We used three
values of $m_l$ to generate the gauge configurations using this
algorithm, with $m_s$ fixed to 0.04 (in lattice units).  Further
parameters and simulation details are given in
Table~\ref{tab:run-details}.

Recently, we developed a more efficient implementation, {RHMC II},
which was used to generate the additional ensemble discussed in
Section~\ref{sec:quotient_RHMC}.  Following Aoki \textit{et al}
\cite{Aoki:2004ht} one may employ a single pseudo-fermion field to
estimate directly each ratio of determinants in
Eq.~(\ref{eq:nf3-det}). This reduces the stochastic noise in the
molecular dynamics evolution and speeds up the calculation because a
larger step size can be used in the integration.
Furthermore, we can rewrite the ratio of fermion determinants as
\begin{widetext}
\begin{equation}
  \det\left\{\frac{{\cal D}(m_l)}{{\cal D}(1)}\right\} \;
  \det\left\{\frac{{\cal D}(m_s)}{{\cal D}(1)}\right\}^{1/2}
  =
  \det\left\{\frac{{\cal D}(m_l)}{{\cal D}(m_s)}\right\} \;
  \det\left\{\frac{{\cal D}(m_s)}{{\cal D}(1)}\right\}^{1/2} \;
  \det\left\{\frac{{\cal D}(m_s)}{{\cal D}(1)}\right\}^{1/2} \;
  \det\left\{\frac{{\cal D}(m_s)}{{\cal D}(1)}\right\}^{1/2},
\label{eq:DeterminantDecomp2}
\end{equation}
\end{widetext}
thus preconditioning the light quark Dirac operator by the strange
quark~\cite{Hasenbusch:2002ai,Urbach:2005ji}. We associate the first determinant
on the RHS of Eq.~(\ref{eq:DeterminantDecomp2}) with a single
pseudo-fermion field, and introduce separate pseudo-fermion fields for
each of the three ``strange quarks'' in the rest of the determinants on the
RHS. Again we used a multiple-timescale integrator, with the
preconditioned light quarks using the largest step size, the
strange/Pauli-Villars fields an intermediate size, and the gauge
fields the finest size. The optimal free parameter of the MN2
integration scheme was found to be $\lambda=0.22$ in this case. The
relevant parameters are also given in Table~\ref{tab:run-details}.

To distinguish the valence quark masses from the sea quark masses, we
will use $m_l^{\rm sea}$ to denote the light dynamical quark masses, and
$m_1$ and $m_2$ to denote the masses of two valence quarks that make
up a meson. Whenever a meson is composed of two degenerate quarks, a
short-hand notation $m_{\rm val}$ may be used to refer to the mass of
each component quark. We use the lattice units for various quantities
throughout the paper, unless physical units are explicitly given. 


\section{Thermalization and Autocorrelations}
\label{sec:ensemble_properties}

\begin{figure}[t]
\centering
\includegraphics[width=\columnwidth]{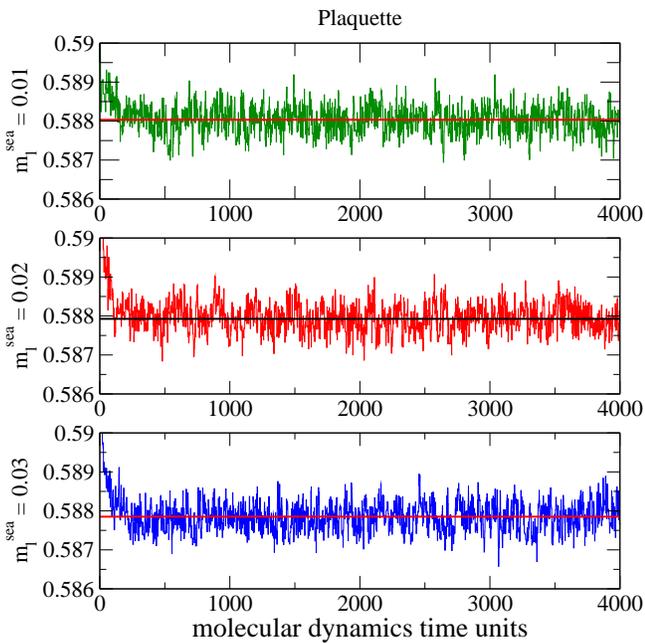}
\caption{Evolution history of the plaquette for all three
  ensembles, measured every MD time unit. The straight lines are the
  averages from  trajectory
  2000 to 4000 for each ensemble.}
\label{fig:plaq_evol}
\end{figure}

\begin{figure}[h]
\centering
\includegraphics[width=\columnwidth]{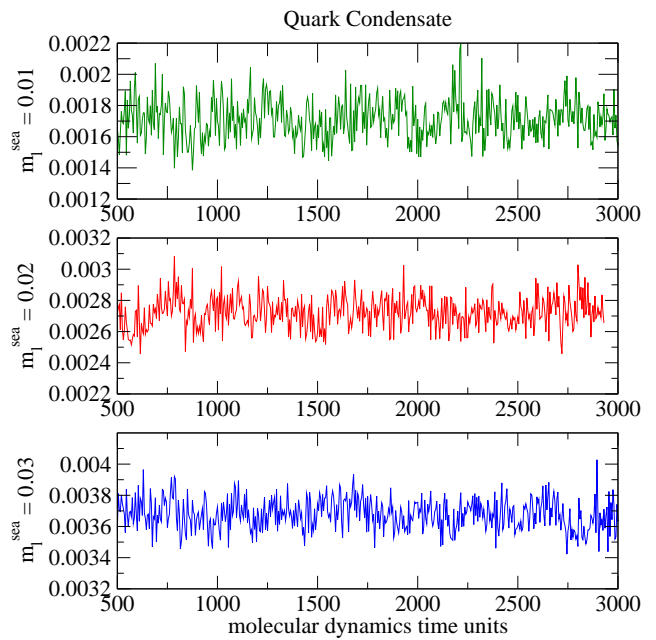}
\caption{Evolution history of the quark condensate
  $\left<\overline{q}q\right>$ for the first 3000 molecular dynamics
  time units at  three dynamical points: $m_{\rm val} =
  m_l^{\rm sea} = 0.01$ (top), $m_{\rm val} = m_l^{\rm sea} = 0.02$
  (middle)  and $m_{\rm val}
  = m_l^{\rm sea} = 0.03$ (bottom). The measurements were taken every 5
  molecular dynamics time units. }
\label{fig:pbp_evol}
\end{figure}

In this section we will discuss the thermalization and
autocorrelations of the ensembles which build a foundation for the
data analysis of various quantities. For most of our measurements, we
discarded the first 500 molecular dynamics time units to account for the thermalization
of the gauge fields.  Shown in Figure~\ref{fig:plaq_evol} is the
evolution history of the plaquette for all three ensembles. The
straight lines are the average values of the plaquette for each
ensemble from time unit 2000 to 4000. It is evident that the gauge fields have come
to equilibrium after ${\cal O}(500)$ molecular dynamics time units. We have also
measured the chiral condensate $\left<\overline{q}q\right>$ through
a stochastic estimator with one single hit for
the first 3000 molecular dynamics time units, starting from time unit 500, and found
no signs of further equilibration (Figure~\ref{fig:pbp_evol}). This
supports our choice  of cut to allow for the thermalization of the
ensemble. 

\begin{figure}[ht]
\centering
\includegraphics[angle=-90,width=\columnwidth]{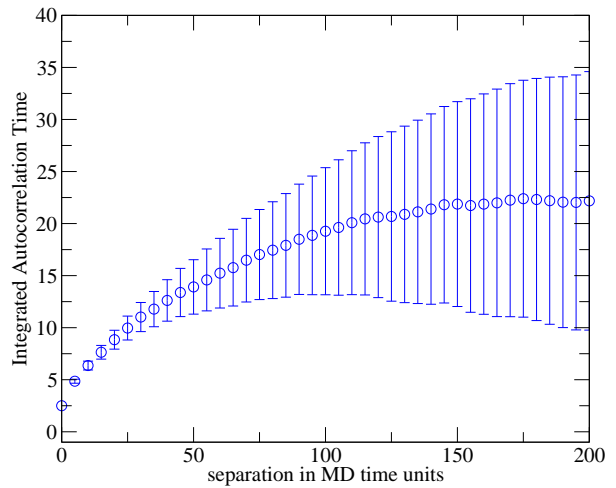}
\caption{Integrated auto-correlation time for the pseudoscalar
  two-point correlation function at time slice 12 with $m_{l}^{\rm sea} =
  m_{\rm val} = 0.01$. }
\label{fig:pscalar_autocorr}
\end{figure}

To obtain reliable estimates for the statistical errors in the physical
observables, we also investigate the autocorrelations in the
ensembles. While the autocorrelation time may differ from quantity to
quantity, we calculate the autocorrelations for the quantities of our
direct interest, such as meson correlation functions. We adopted the same
method as in \cite{Aoki:2004ht} and calculated the integrated
autocorrelation time, $\tau_{\rm int}$, for the two-point pseudoscalar
correlation function at time slice 12 obtained from a Coulomb gauge
fixed wall  source at $t=0$ with two degenerate valence quarks with
$m_{\rm val}= 0.01$  for the $m_l^{\rm sea} = 0.01$ ensemble, where the
correlation function was measured every 5 molecular dynamics time
units. This is shown in  Figure~\ref{fig:pscalar_autocorr}. We can see
that the integrated autocorrelation time reaches a plateau of about 20
to 25 time units. The separation between two independent measurements
is then  about $2\tau_{\rm int}
\approx 50$ molecular dynamics time units. Thus we chose to average the
data over a block of measurements in such a way that the span of the
measurements in each block covers at least 50 molecular dynamics time
units whenever possible prior to the standard jackknife analysis. 

\begin{figure}[ht]
\centering
\includegraphics[angle=-90,width=\columnwidth]{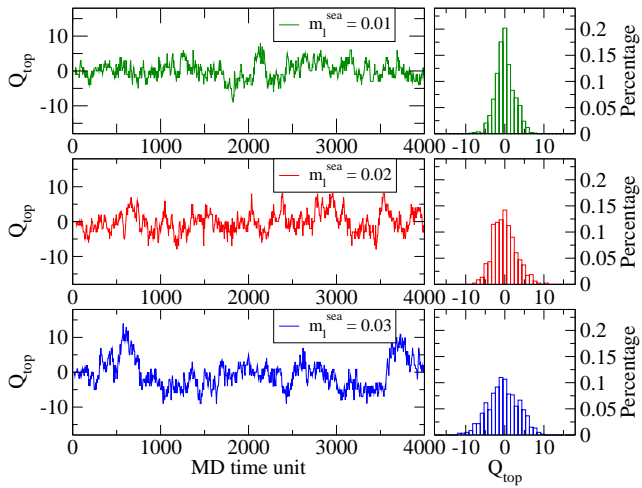}
\caption{The topological charge as measured by
  gluonic fields on the three ensembles. The left three panels are the
  topological charge  as a function of the molecular dynamics
  time. Each of the three  graphs on the right represents the
  normalized  histogram of the corresponding topological charge evolution on the left.  }
\label{fig:topology}
\end{figure}

The evolution of the topological charge is another indicator of how
fast the gauge fields  decorrelate. On each ensemble we measured the
topological  charge using the classically ${\cal O}(a^2)$ improved
definition of the field tensor as defined in
\cite{Antonio:2006px}. Figure~\ref{fig:topology} shows that the gauge
field  moves
frequently between different topological charge sectors. Also shown in
Figure~\ref{fig:topology} are the histograms of the topological
charge, which  indicate that different topological sectors are sampled
reasonably  well. However, it is also worth noting
that correlations on a scale of a few hundred molecular dynamics time
units are present
in the $m_l^{\rm sea} = 0.03$ ensemble. This may imply that the value for 
the autocorrelation time discussed in the previous paragraph is under-estimated. 
Ideally we should use a larger block size prior to the
jackknife analysis to have more robust error estimates, but we are
constrained by the moderate lengths of the simulations and have chosen
to use  a block
size that is not too small, but at the same time leaves a reasonable
number of jackknife blocks to perform covariant fits when necessary. 

\section{Hadronic Observables}
\label{sec:observables}
\subsection{Calculation Details and Fitting Procedures}
\label{sec:fitting}

We used several different source and sink operators to calculate the
quark  propagators needed for the two-point correlation functions. To
be specific,  we used a
local operator (denoted as L), a Coulomb gauge fixed wall operator
(denoted as W),  and a Gaussian smeared
operator~\cite{Allton:1993wc} (denoted as G), for the source and/or
sink, in an attempt to
investigate the systematic uncertainties coming from excited-state
contamination. Here we adopt the same notation as
in~\cite{Antonio:2006px} and  denote the types of meson correlation
functions by the different source and sink operators used in the quark
propagators. For example, GL-GL means a correlation function
composed of two quark propagators, each of which is computed using the
Gaussian  smeared source operator and the local sink operator.  The
sources are placed at multiple locations to reduce the short-term
noise induced by the fluctuations in the gauge fields. The correlation
functions are averaged over different source locations before the standard
jackknife analysis is performed.  Details of the calculation
parameters are tabulated in Tables~\ref{tab:meas_WL} and
\ref{tab:meas_GL}.

\begin{table}[b]
\caption{Measurement parameters for the WL-WL correlation
  functions for all the three ensembles. $m_l^{\rm sea}$ denotes the
  light sea quark mass. $m_1$ and $m_2$ denote the two valence quarks
  composing the mesons. $\Delta$ is the separation between
  measurements in  molecular dynamics time units.} \label{tab:meas_WL}
\begin{tabular*}{\columnwidth}{@{\extracolsep{\fill}}ccccccc}
\hline
\hline
$m_l^{\rm sea}$ & $m_1$ & $m_2$ & range & $\Delta$ & $N_{\rm meas}$ & $t_{\rm src}$
\\
\hline
0.01 & 0.01 & 0.01 & 500 - 3995 & 5 & 700 & 0, 16 \\ 
0.01 & 0.02 & 0.02 & 500 - 3995 & 5 & 700 & 0, 16 \\ 
0.01 & 0.03 & 0.03 & 500 - 3995 & 5 & 700 & 0, 16 \\ 
0.01 & 0.04 & 0.04 & 500 - 3995 & 5 & 700 & 0, 16 \\
\hline
0.02 & 0.01 & 0.01 & 500 - 4045 & 5 & 710 & 0, 16 \\ 
0.02 & 0.02 & 0.02 & 500 - 4045 & 5 & 710 & 0, 16 \\ 
0.02 & 0.03 & 0.03 & 500 - 4045 & 5 & 710 & 0, 16 \\ 
0.02 & 0.04 & 0.04 & 500 - 4045 & 5 & 710 & 0, 16 \\
\hline
0.03 & 0.01 & 0.01 & 500 - 3995 & 5 & 700 & 0, 16 \\ 
0.03 & 0.02 & 0.02 & 500 - 3995 & 5 & 700 & 0, 16 \\ 
0.03 & 0.03 & 0.03 & 500 - 3995 & 5 & 700 & 0, 16 \\ 
0.03 & 0.04 & 0.04 & 500 - 3995 & 5 & 700 & 0, 16 \\
\hline
\hline
\end{tabular*}
\end{table}

\begin{table}[ht]
\caption{Measurement parameters for the LL-LL, GG-GG and GL-GL correlation
  functions for all the three ensembles. $m_l^{\rm sea}$ denotes the
  light sea quark mass. $m_1$ and $m_2$ denote the two valence quarks
  composing the mesons. $\Delta$ is the separation between
  measurements in  molecular dynamics time units. } \label{tab:meas_GL}
\begin{tabular*}{\columnwidth}{@{\extracolsep{\fill}}ccccccc}
\hline
\hline
$m_l^{\rm sea}$ & $m_1$ & $m_2$ & range & $\Delta$ & $N_{\rm meas}$ & $t_{\rm src}$
\\
\hline
\hline
0.01 & 0.01 & 0.01 & 500 - 3990 & 10 & 350 & 0, 16 \\ 
0.01 & 0.01 & 0.04 & 500 - 3990 & 10 & 350 & 0, 16 \\ 
0.01 & 0.04 & 0.04 & 500 - 3990 & 10 & 350 & 0, 16 \\
0.01 & 0.01 & 0.01 & 505 - 3995 & 10 & 350 & 8, 24 \\ 
0.01 & 0.01 & 0.04 & 505 - 3995 & 10 & 350 & 8, 24 \\ 
0.01 & 0.04 & 0.04 & 505 - 3995 & 10 & 350 & 8, 24 \\
\hline
0.02 & 0.02 & 0.02 & 500 - 3990 & 10 & 350 & 0, 16 \\ 
0.02 & 0.02 & 0.04 & 500 - 3990 & 10 & 350 & 0, 16 \\ 
0.02 & 0.04 & 0.04 & 500 - 3990 & 10 & 350 & 0, 16 \\
0.02 & 0.02 & 0.02 & 505 - 3995 & 10 & 350 & 8, 24 \\ 
0.02 & 0.02 & 0.04 & 505 - 3995 & 10 & 350 & 8, 24 \\ 
0.02 & 0.04 & 0.04 & 505 - 3995 & 10 & 350 & 8, 24 \\

\hline
0.03 & 0.03 & 0.03 & 500 - 3990 & 10 & 350 & 0, 16 \\ 
0.03 & 0.03 & 0.04 & 500 - 3990 & 10 & 350 & 0, 16 \\ 
0.03 & 0.04 & 0.04 & 500 - 3990 & 10 & 350 & 0, 16 \\
\hline
\hline
\end{tabular*}
\end{table}

In parallel to the spectrum calculations, we have also done another set of
measurements
necessary for the extraction of weak matrix elements (see, for
example, Ref.~\cite{Blum:2001xb}), in which the quark
propagators were calculated from Coulomb gauge
fixed wall sources at time slices $t_{\rm src} = 5$ and $t_{\rm src} =
27$. In contrast to the standard spectrum measurement, here the meson
correlation functions were constructed from the sum of
propagators computed with periodic (P) and anti-periodic (A) boundary
conditions in the time direction. This has the effect of doubling the temporal
extent
of the lattice which gives a longer time range
to extract the meson masses. Five different valence masses
are used to compute the quark propagators and the meson correlation
functions are constructed from all possible combinations of
these propagators. The details of measurements are
given in Table~\ref{tab:meas_3pt}. We will use ``P+A'' to refer to the
results obtained from these data sets hereafter.

\begin{table}[t]
\centering
\caption{Measurement parameters for the correlation
  functions measured with P+A boundary conditions for all the three
  ensembles. $m_l^{\rm sea}$  denotes the
  light sea quark mass. $m_1$ and $m_2$ denote the two valence quarks
  composing the mesons. $\Delta$ is the separation between
  measurements in molecular dynamics time units.} \label{tab:meas_3pt}
\begin{tabular*}{\columnwidth}{@{\extracolsep{\fill}}ccccccc}
\hline
\hline
$m_l^{\rm sea}$ & $m_1$ & $m_2$ & range & $\Delta$ & $N_{\rm meas}$ & $t_{\rm src}$
 \\
\hline
0.01 & 0.01 - 0.05 & 0.01 - 0.05 & 1000 - 4000 & 20 & 150  & 5, 27  \\ 
\hline
0.02 & 0.01 - 0.05 & 0.01 - 0.05 & 1000 - 4000 & 20 & 150  & 5, 27  \\ 
\hline
0.03 & 0.01 - 0.05 & 0.01 - 0.05 & 1000 - 4000 & 20 & 150  & 5, 27  \\ 
\hline
\hline
\end{tabular*}
\end{table}

To get the meson masses and the corresponding amplitudes of the
correlation functions,  we  performed
the standard covariant $\chi^2$ fits to the correlation functions using the
hyperbolic cosine form:
\begin{equation}
  C(t) = A \left [ e^{-m_Gt} + e^{-m_G(T-t)} \right ]
\label{eq:cosh}
\end{equation}
where $m_G$ is the ground-state meson mass, and $t$ is the time slice
relative to the  source. Note that here $T$ is the extent in the
time direction of the lattice, {\it i.e.} 32 for the standard spectrum
measurements and 64 for
the P+A measurements. This relation assumes
that the separation between the source and sink operators is large
enough that the contamination from the excited states is
negligible. However, different source operators have different degrees
of overlap with excited states, and hence the minimum 
time slice that should be included in each fit may differ. To reduce
the uncertainties in fit parameters from the choice of fit range, we
also performed simultaneous fits to two types of correlation functions
using a double-cosh form which give both the ground-state and the
first-excited-state meson masses. 
We found that the ground-state meson masses obtained from the
double-cosh fits were  consistent with the simple cosh fit in
Eq.(\ref{eq:cosh}). Since  we are mostly interested in the
ground-state meson masses in  this paper, we will only present results
obtained from the  single cosh fits unless otherwise stated. 

\begin{figure}[htb]
\centering
\begin{tabular}{c}
\includegraphics[angle=-90,width=\columnwidth]{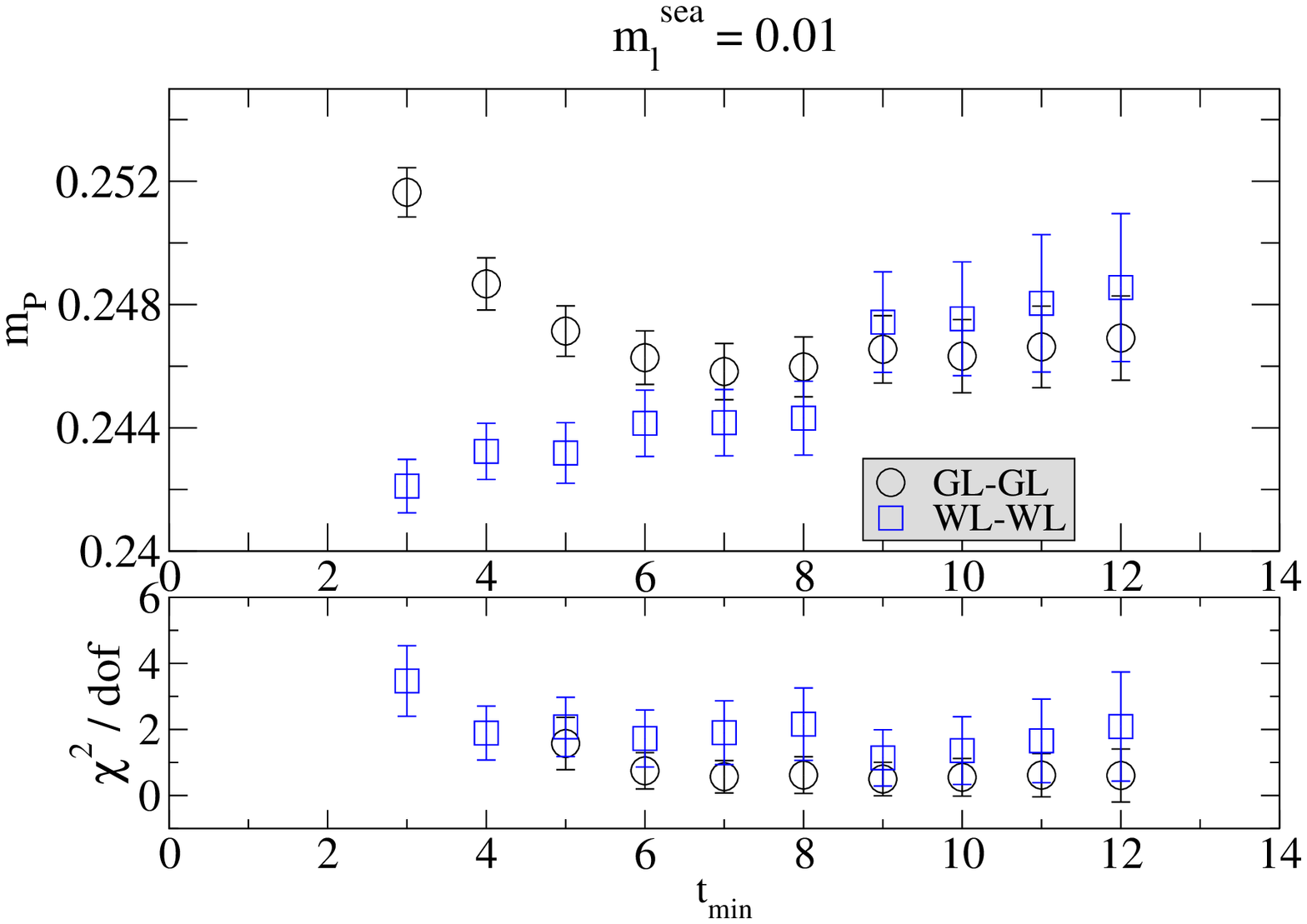} \\
\includegraphics[angle=-90,width=\columnwidth]{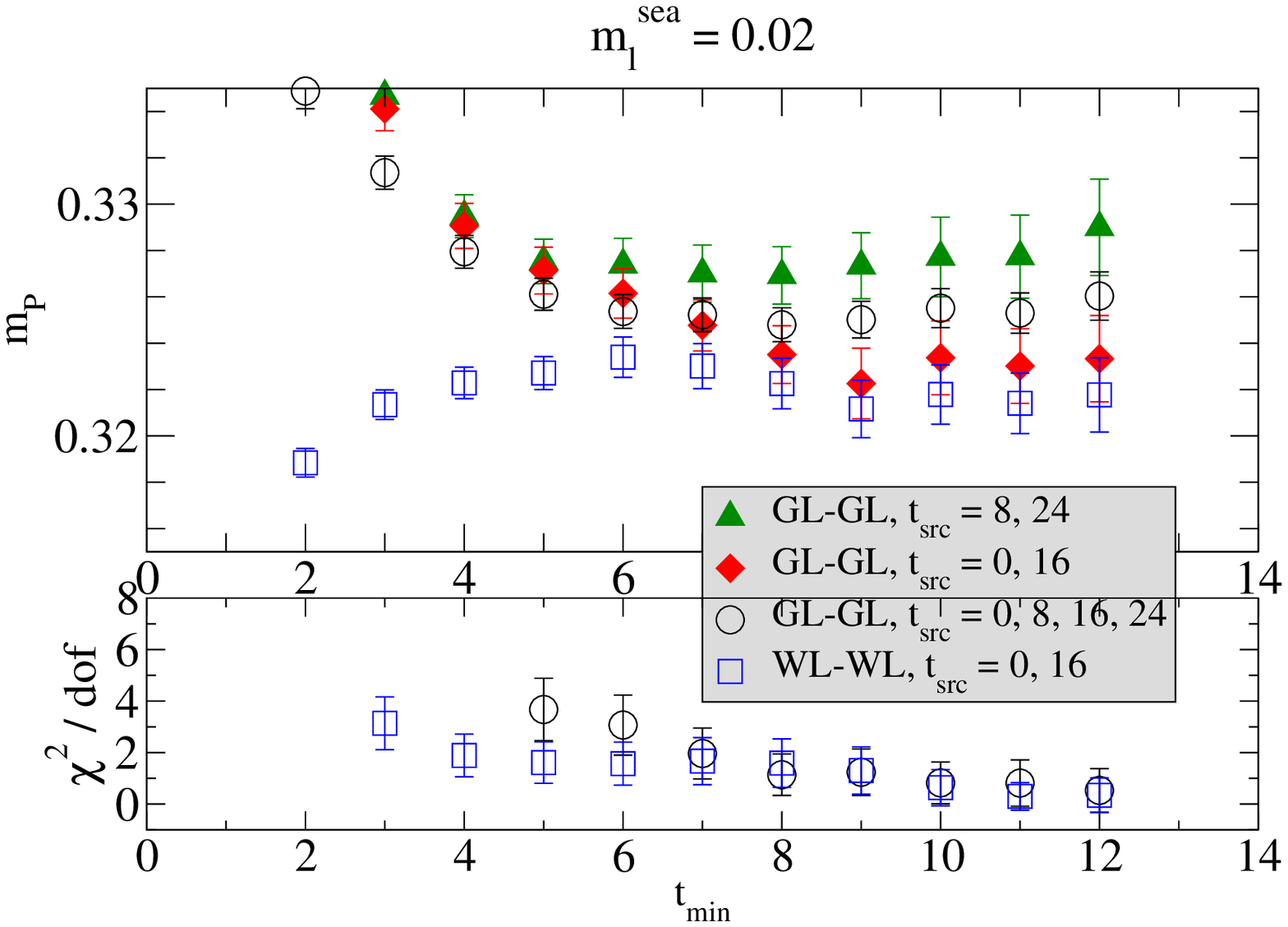} \\
\includegraphics[angle=-90,width=\columnwidth]{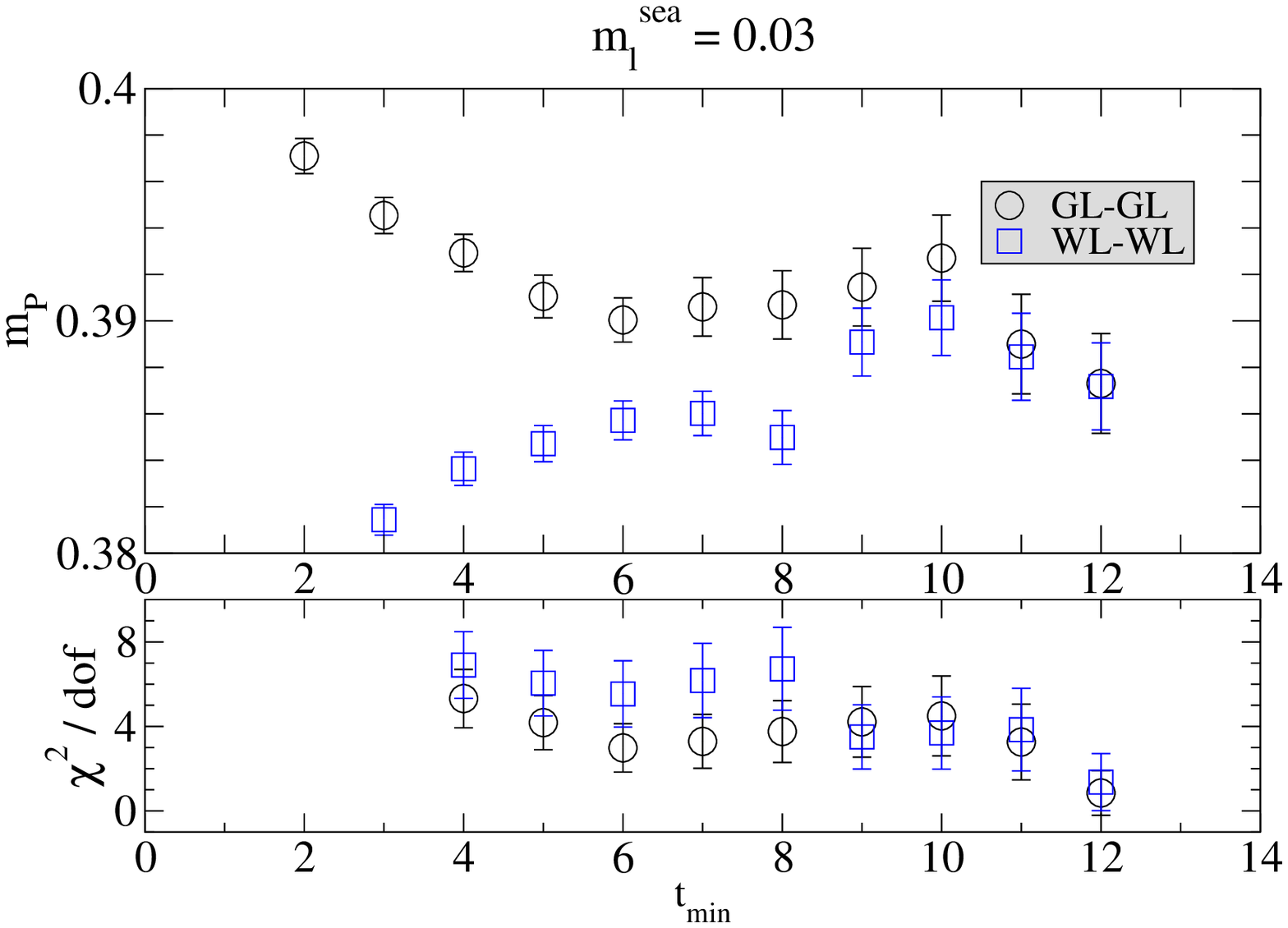} \\
\end{tabular}
\caption{Variation of the pseudoscalar masses obtained from the correlated fits and
  the corresponding $\chi^2$/dof with respect to $t_{\rm min}$ for
  each of the  dynamical points.}
\label{fig:pseudo_fit}
\end{figure}

In order to determine a proper fit range for the meson masses using
Eq.(\ref{eq:cosh}),  we examined the
effective masses as defined by
\begin{equation}
m_{\rm eff} = {\rm cosh}^{-1}\left [ \frac{C(t+1) + C(t-1)}{2 C(t)} \right]
\end{equation}
and chose the minimum time slice in the fit to be the onset of the
plateau in  the
effective mass. A more stringent way to determine the best fit range
is to check  how
the mass obtained from the fit and the resulting $\chi^2$/dof vary
with the fit range. Since the fits are insensitive to the maximum
time slice used, we only investigated the variations with respect to
the lower bound (denoted as $t_{\rm min}$) and fixed the upper bound
(denoted as $t_{\rm max}$) to $T/2$ for the GL-GL and WL-WL data
sets. In  Figure
\ref{fig:pseudo_fit}, the fitted masses and $\chi^2$/dof
for the pseudoscalar correlation functions are plotted for each light
dynamical point with $m_{\rm val}=m_l^{\rm sea}$. As we can see, the
central  values of the fitted masses are
insensitive to
the choice of $t_{\rm min}$ when $t_{\rm min}$ is above 8 or 9 for
both GL-GL and WL-WL data sets. To obtain high confidence level for
the fits, we also seek to have minimal $\chi^2$/dof in
the time range where the fits are stable. 

Typically the fitted pseudoscalar meson masses from WL-WL and GL-GL correlation
functions
agree quite well when $t_{\rm min}$ is above 9. An exception is the $m_l^{\rm sea}
= 0.02$ data set, where results from
GL-GL correlation functions are larger than those from WL-WL
correlation functions by a few standard deviations. The deviations
result from the fact that the GL-GL
measurements  were performed at an additional set of source locations
which were not used in the WL-WL measurements.  When only the data with the
source operator at
time slices 0 and 16 are included in the analysis, the results
obtained from GL-GL and WL-WL agree within statistical
uncertainties, whilst the GL-GL measurements from sources at time
slices 8 and 24 give much larger values. Hence when all the sources are
combined in the analysis, the values of the fitted
masses are much higher from the GL-GL correlators than those from the
WL-WL correlators.  The fact
that the results are so sensitive to the different
source locations indicates that we have not sampled phase space
extremely well.  Thus, the errors obtained from the standard jackknife
procedure may  be under-estimated. 

%
%

\subsection{Meson Masses}
\label{sec:meson_mass}

\begin{figure}[t]
\centering
\includegraphics[width=\columnwidth]{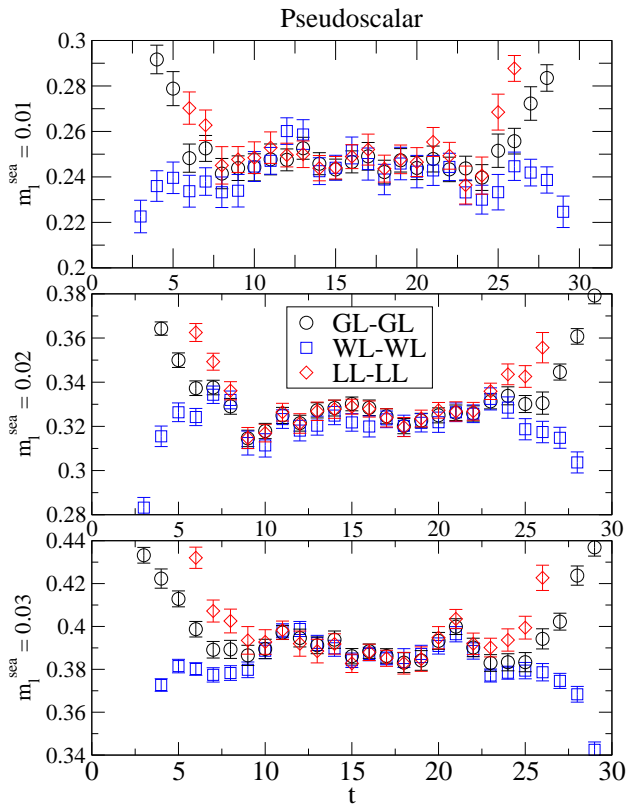}
\caption{Pseudoscalar effective masses from WL-WL, LL-LL and GL-GL
  correlation functions for the three ensembles. The valence quarks
  are degenerate, with the masses equal to the light sea quark mass
  of that ensemble.}
\label{fig:pseudo_effm_comp}
\end{figure}

The pseudoscalar meson masses can be obtained from several different
correlation functions \cite{Blum:2000kn,Antonio:2006px}. We 
will show the results from both the
pseudoscalar (PP) and axial vector (AA)
channels.  The  effective masses in the
pseudoscalar channel at the light dynamical points for all three
ensembles are shown in Figure~\ref{fig:pseudo_effm_comp}.  The
effective masses  from LL-LL correlation functions typically have
later approaches  to plateaus and leave us with fewer time slices to
perform the  fits. Thus we will not quote results from these data sets
unless  otherwise noted. We also note that there
are some variations in the effective masses greater than estimates of
the standard deviation from the
variance of the data. The presence of these variations
 with 4000 MD time units may be another indication of long-term
 autocorrelations in these ensembles. Although we do not fully
 understand the  cause of this problem, it is likely due to
the algorithm we used to generate these gauge configurations. This
will be further  addressed in Section~\ref{sec:quotient_RHMC}.    

\begin{figure}[htp]
\centering
\includegraphics[width=\columnwidth]{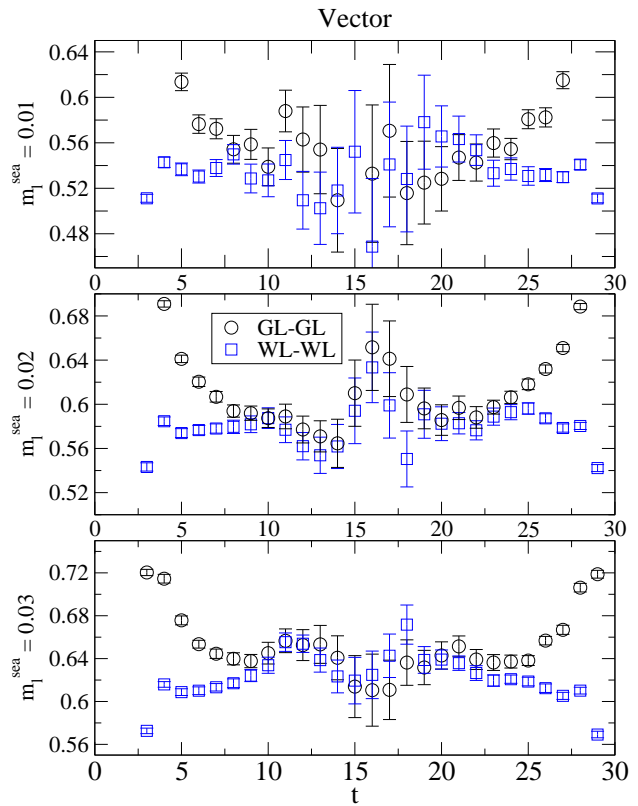}
\caption{Vector effective masses from WL-WL and GL-GL
  correlation functions for the three ensembles. The valence quarks
  are degenerate, with the masses equal to the light sea quark mass
  of that ensemble.}
\label{fig:vector_effm}
\end{figure}

We determined the vector meson masses by averaging over the three
polarizations to reduce statistical fluctuations. Some representative
effective masses for the vector mesons are shown in
Figure~\ref{fig:vector_effm}. They are typically noisier than the
pseudoscalar effective masses, and approach plateaus later. As the
quark masses become lighter, the noise on the effective masses increases. 

%
%
\begin{table}[htbp]
\centering
\caption{Results of the pseudoscalar and vector  meson masses averaged
  from different  measurements. The quoted errors are the largest
  statistical errors  of the different measurements multiplied by 1.5
  to account for  possible underestimation of errors as described in the text. }
\label{tab:summary_table}
\begin{tabular*}{\columnwidth}{@{\extracolsep{\fill}}ccccc}
\hline
\hline
$m_l^{\rm sea}$ & $m_{1}$  & $m_2$ & $m_{P}$ &  $m_{V}$\\
\hline
0.01 & 0.01 & 0.01 & 0.247(3) & 0.549(20) \\
0.01 & 0.02 & 0.01 & 0.290(3) & 0.564(14) \\
0.01 & 0.02 & 0.02 & 0.323(3) & 0.577(11) \\
0.01 & 0.03 & 0.01 & 0.325(3) & 0.580(11) \\
0.01 & 0.03 & 0.02 & 0.357(3) & 0.595(9)   \\
0.01 & 0.03 & 0.03 & 0.385(3) & 0.609(8)   \\
0.01 & 0.04 & 0.01 & 0.356(3) & 0.599(10) \\
0.01 & 0.04 & 0.02 & 0.387(3) & 0.611(8)   \\
0.01 & 0.04 & 0.03 & 0.414(3) & 0.626(7)   \\
0.01 & 0.04 & 0.04 & 0.438(3) & 0.642(6)   \\
0.01 & 0.05 & 0.01 & 0.387(3) & 0.613(9)   \\
0.01 & 0.05 & 0.02 & 0.414(3) & 0.627(7)   \\
0.01 & 0.05 & 0.03 & 0.440(3) & 0.642(6)   \\
0.01 & 0.05 & 0.04 & 0.465(3) & 0.657(6)   \\
0.01 & 0.05 & 0.05 & 0.489(3) & 0.672(5)   \\
\hline
0.02 & 0.01 & 0.01 &  0.250(3) & 0.546(20) \\
0.02 & 0.02 & 0.01 &  0.292(3) & 0.560(14) \\
0.02 & 0.02 & 0.02 &  0.325(3) & 0.585(11) \\
0.02 & 0.03 & 0.01 &  0.327(3) & 0.579(12) \\
0.02 & 0.03 & 0.02 &  0.358(3) & 0.597(9) \\
0.02 & 0.03 & 0.03 &  0.386(3) & 0.615(8) \\
0.02 & 0.04 & 0.01 &  0.359(3) & 0.597(10)\\
0.02 & 0.04 & 0.02 &  0.388(3) & 0.618(8) \\
0.02 & 0.04 & 0.03 &  0.415(3) & 0.631(7) \\
0.02 & 0.04 & 0.04 &  0.440(3) & 0.649(6) \\
0.02 & 0.05 & 0.01 &  0.388(3) & 0.614(9) \\
0.02 & 0.05 & 0.02 &  0.415(3) & 0.631(7) \\
0.02 & 0.05 & 0.03 &  0.441(2) & 0.647(6) \\
0.02 & 0.05 & 0.04 &  0.466(2) & 0.663(6) \\
0.02 & 0.05 & 0.05 &  0.490(2) & 0.679(5) \\
\hline
0.03 & 0.01 & 0.01 &  0.251(3) & 0.599(24) \\
0.03 & 0.02 & 0.01 &  0.289(3) & 0.589(17) \\
0.03 & 0.02 & 0.02 &  0.325(3) & 0.613(12) \\
0.03 & 0.03 & 0.01 &  0.324(3) & 0.603(13) \\
0.03 & 0.03 & 0.02 &  0.355(3) & 0.616(10) \\
0.03 & 0.03 & 0.03 &  0.387(3) & 0.643(9) \\
0.03 & 0.04 & 0.01 &  0.356(3) & 0.618(11) \\
0.03 & 0.04 & 0.02 &  0.385(3) & 0.631(9) \\
0.03 & 0.04 & 0.03 &  0.412(3) & 0.655(8) \\
0.03 & 0.04 & 0.04 &  0.442(3) & 0.670(7) \\
0.03 & 0.05 & 0.01 &  0.386(3) & 0.633(10) \\
0.03 & 0.05 & 0.02 &  0.413(3) & 0.646(8) \\
0.03 & 0.05 & 0.03 &  0.438(3) & 0.660(7) \\
0.03 & 0.05 & 0.04 &  0.463(3) & 0.675(6) \\
0.03 & 0.05 & 0.05 &  0.488(2) & 0.690(6) \\
\hline
\hline
\end{tabular*}
\end{table}
As discussed already, we have chosen to fit the correlation functions
with different source/sink operators independently to obtain the
corresponding meson masses, which are given in Appendix \ref{sec:appendix_tables}.
The sources differ not only in their spatial characteristics, but
also in their temporal location in the lattices.  As such, they
sample different parts of the gauge fields.  The different combinations
of source/sink operators give masses which generally agree, but
there are a number of cases where the difference is somewhat outside
of the error bars.  Since we are concerned about effectively sampling
phase space, the quoted statistical errors may be under-estimated
due to possible long-range autocorrelations.  We can use the fact
that the different sources sample different parts of the lattice to
compensate for this.  For example, for the $m_l^{\rm sea} = 0.02$ ensemble,
we find the $m_1 = m_2 = 0.02$ pseudoscalar meson mass has a value of 0.3247(8),
0.3212(12) and 0.3251(18) from the PP channel, and 0.3254(12),
0.3226(11) and 0.3287(19) from the AA channel.  Taking these six
values as  independent
and calculating a standard deviation from them gives 0.3246(26),
which has an error bar larger than from any of the individual
measurements.  From this example, amongst the ones with the largest
differences, we have chosen to take the averages of the results for
the meson masses obtained from different measurements, wherever
possible, as our best estimates of the meson masses.  Since the
different measurements involve different source locations on our
lattices, the spread of values is also an indication of the statistical
errors of our ensembles.  We correspondingly inflate the largest
individual error by a factor of 1.5 to bring it in rough agreement
with this spread.  The final results are summarized in
Table~\ref{tab:summary_table}, where the values of $m_{P}$ are taken
from the averages of both the PP and AA channels
in different measurements.  These mass values will then be used in
the subsequent discussions of the paper.

%
%
\subsection{Residual Mass}
\label{sec:mres}

\begin{figure}[t]
\centering
\includegraphics[width=\columnwidth]{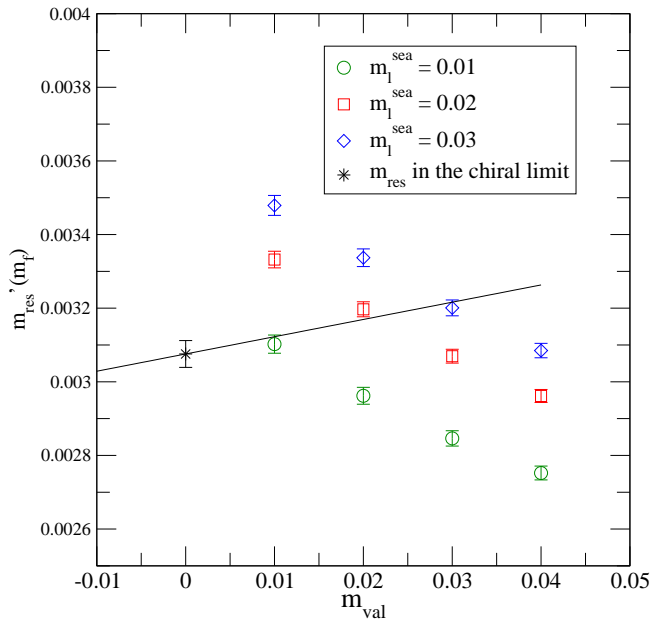}
\caption{The residual mass determined from the WL-WL correlators. The
  solid line is a linear  fit to the dynamical points with $m_{\rm val} = m_l^{\rm sea}$.}
\label{fig:mres}
\end{figure}

The quantitative description of the explicit chiral symmetry breaking
for domain  wall fermions is the residual mass,
$\mres$~\cite{Blum:2000kn,Antonio:2006px},  which is determined from the ratio
\begin{equation}
  R(t) =
  \frac{\langle \sum_{\bm{x}} J^a_{5q} (\bm{x}, t)
  \pi^a(0) \rangle }
  {\langle \sum_{\bm{x}} J^a_5(\bm{x}, t)
  \pi^a(0) \rangle }.
  \label{eq:mres_ratio}
\end{equation}
The definitions of $J^a_{5q}(\bm{x},t)$ and $J^a_{5}(\bm{x},t)$ for
domain wall fermions can be found in Ref.~\cite{Blum:2000kn}. We
follow ~\cite{Antonio:2006px}  and determine $\mres$ at a given quark
mass, denoted by $\mres '(m_f)$, by fitting $R(t)$ to a constant over
a time range where $R(t)$ shows a good plateau. The results from the
WL-WL correlation functions are shown in
Table~\ref{tab:mres_WL}.

While by definition $\mres$ should be a constant, the operator we
choose to determine it happens to depend on the quark mass, as shown in
Figure~\ref{fig:mres}.  This quark mass dependence comes from the
higher-dimension operators in the divergence of the axial current, which should
be suppressed by powers  of
$a$~\cite{Antonio:2006px,Lin:2006cf,Blum:2001sr}. We choose to deal 
with these lattice artifacts by extrapolating $\mres '(m_f)$ to 
the zero quark mass limit in both the sea and valence sectors. Doing so gives
\begin{equation}
\mres = 0.00308(4). 
\end{equation}
This is about 1/3 of the lightest quark mass in our simulations, which
may seem larger than ideal. Nevertheless, from a field
theoretic point of view, to leading order the residual mass is just an additive
renormalization to the input quark mass. Neglecting higher-order
corrections, the chiral limit is defined at $m_f + \mres = 0$ or $m_f
= -\mres$, where $m_f$ is the input quark mass. This is the basis for
the chiral extrapolations discussed in Section~\ref{sec:chiral_limit}. In
this paper we
do not attempt to discuss higher-order corrections to the residual
mass, which should be small in our simulations.

\begin{table}[t]
\centering
\caption{Residual mass  obtained with the Coulomb gauge fixed wall source. }
\label{tab:mres_WL}
\begin{tabular*}{\columnwidth}{@{\extracolsep{\fill}}cccccc}
\hline
\hline
$m_l^{\rm sea}$ & $m_{1}$  & $m_2$ & $t_{\rm min}-t_{\rm max}$ &
$\chi^2/dof$ & $\mres '(m_f)$ \\
\hline
0.01 & 0.01 & 0.01 & 8-16 &  0.9(7)  & 0.003102(25) \\
0.01 & 0.02 & 0.02 & 8-16 &  1.0(7)  & 0.002962(23) \\
0.01 & 0.03 & 0.03 & 8-16 &  1.0(7)  & 0.002846(21) \\
0.01 & 0.04 & 0.04 & 8-16 &  0.9(7)  & 0.002752(19) \\
\hline
0.02 & 0.01 & 0.01 & 8-16 &  1.0(7)  & 0.003332(23) \\
0.02 & 0.02 & 0.02 & 8-16 &  0.9(7)  & 0.003197(20) \\
0.02 & 0.03 & 0.03 & 8-16 &  0.9(7)  & 0.003070(19) \\
0.02 & 0.04 & 0.04 & 8-16 &  0.9(7)  & 0.002961(18) \\
\hline
0.03 & 0.01 & 0.01 & 8-16 &  1.3(8)  & 0.003479(27) \\
0.03 & 0.02 & 0.02 & 8-16 &  1.5(9)  & 0.003337(24) \\
0.03 & 0.03 & 0.03 & 8-16 &  1.7(9)  & 0.003201(22) \\
0.03 & 0.04 & 0.04 & 8-16 &  1.9(10)  & 0.003085(20) \\
\hline
\hline
\end{tabular*}
\end{table}

%
%
\subsection{Pseudoscalar Decay Constants}
\label{sec:fpi}

 The decay constant $f_{P}$ for a charged pseudoscalar meson is defined by 
\begin{equation}
-i\,f_{P}\, q_\mu e^{-iq\cdot x} = { \langle\, 0\, | \,{\cal A}_\mu(x)\, |\, \pi (\bm{q}) \,\rangle},
\label{eq:fpi}
\end{equation}
where ${\cal A}_\mu(x)$ is the (partially) conserved axial vector
current operator for  domain wall fermions~\cite{Blum:2000kn}, and $|\pi(\bm{q})\rangle$ 
is the state vector for a pseudoscalar meson with momentum $\bm{q}$. We normalize the state vector such that $\langle \pi(\bm{q}) | \pi(\bm{q^\prime}) \rangle = (2\pi)^3 
\,2\, E_{\bm{q}}\, \delta(\bm{q}-\bm{q^\prime})$. ${\cal A}_\mu$ is
related to the local  axial vector operator by ${\cal A}_\mu = Z_A A_\mu$
for  physics at long  distances. It is conventional to determine the
matrix element of  $\langle 0 | { A}_0 | \pi\rangle$ from the
amplitude of the LL-LL  correlation functions~\cite{Antonio:2006px,Blum:2000kn},
$A_{A_0,A_0}^{LL}$. Here a double-cosh fit to  the LL-LL and GG-GG
correlation functions was  performed to get a more reliable estimate
of the amplitude of the  LL-LL correlation functions. $f_{P}$ is obtained by 
\begin{equation}
 f^{(1)}_{P} = Z_A \sqrt{\frac{2 A^{LL}_{A_0, A_0}}{m_{P}}}.
\label{eq:fpi_LL}
\end{equation}

 A novel way of determining $f_{P}$ is from the WL-WL correlators. In
 contrast to the point-like  sources,  the normalization of the
 Coulomb gauge fixed wall source  is not known analytically, and we
 need to compute the relative  normalization of the source operator to
 the conserved current  numerically. This can be done by employing the
 following ratio, 
 \begin{equation}
  N_W  =  \frac{\sum_{\bm{x}}\langle {\cal A}_0 (\bm{x},t) P^W(0)
  \rangle }{\sum_{\bm{x}}\langle   A_0^W (\bm{x},t) P^W(0) \rangle},
\end{equation}
where $P^W(0)$ is the Coulomb gauge fixed pseudoscalar source operator. 
The pseudoscalar decay constant can then be computed using the following formula:
\begin{equation}
  f^{(2)}_{P} = \sqrt{  \frac{2Z_A N_W A_{A_0,A_0}^{WL}}{m_{P}}}.
\label{eq:fpi_WL}
\end{equation}
This method has the advantage of giving smaller statistical errors
compared to the conventional  way of determining $f_{P}$ from the
point-like  sources.
We note that here $N_W$ is merely a normalization factor for the two
operators  ${\cal A}_0 (\bm{x},t)$  and $A_0^W (\bm{x},t)$ computed
under the same condition. Unlike $Z_A$, which is a constant up to
corrections of ${\cal O}(a^2)$, the mass dependence of $N_W$ has
physical significance and thus it is the value of $N_W$ evaluated at
each valence quark mass that should be used in Eq.(\ref{eq:fpi_WL}). 
\begin{table}[t]
\centering
\caption{$Z_A$  from WL-WL correlation functions
  for all three ensembles.}
\label{tab:Za_WL}
\begin{tabular*}{\columnwidth}{@{\extracolsep{\fill}}cccccc}
\hline
\hline
$m_l^{\rm sea}$ & $m_{1}$  & $m_2$ & $t_{\rm min}-t_{\rm max}$ & $\chi^2/dof$ & $Z_A$ \\
\hline
0.01 & 0.01 & 0.01 & 8-16 &  0.5(5)  & 0.71807(14) \\
0.01 & 0.02 & 0.02 & 8-16 &  0.7(6)  & 0.71930(11) \\
0.01 & 0.03 & 0.03 & 8-16 &  0.8(6)  & 0.72071(9) \\
0.01 & 0.04 & 0.04 & 8-16 &  0.9(7)  & 0.72222(8) \\
\hline
0.02 & 0.01 & 0.01 & 8-16 &  1.2(8)  & 0.71799(14) \\
0.02 & 0.02 & 0.02 & 8-16 &  1.0(7)  & 0.71925(10) \\
0.02 & 0.03 & 0.03 & 8-16 &  0.9(7)  & 0.72069(8) \\
0.02 & 0.04 & 0.04 & 8-16 &  0.8(7)  & 0.72221(7) \\
\hline
0.03 & 0.01 & 0.01 & 8-16 &  1.7(9)  & 0.71806(15) \\
0.03 & 0.02 & 0.02 & 8-16 &  2.0(10)  & 0.71931(11) \\
0.03 & 0.03 & 0.03 & 8-16 &  2.2(11)  & 0.72073(10) \\
0.03 & 0.04 & 0.04 & 8-16 &  2.2(10)  & 0.72226(9) \\
\hline
\hline
\end{tabular*}
\end{table}

The value of $Z_A$ is determined using the method described in
\cite{Antonio:2006px} and \cite{Blum:2000kn}. We quote the results
obtained from  the WL-WL
functions in Table~\ref{tab:Za_WL}. Results from other source-sink
combinations are in good agreement with the shown values, although we
do not display them in the paper. Taking the chiral limit removes
an ${\cal O}(a^2)$ lattice artifact and gives 
\begin{equation}
Z_A = 0.7162(2).
\end{equation}
This is then used to extract $f_{P}$ from the axial vector
correlators using either Eq.(\ref{eq:fpi_LL}) or Eq.(\ref{eq:fpi_WL}),
the results of which are given in Tables~\ref{tab:fpi_GL} and
\ref{tab:fpi_WL}, respectively. These two methods of extracting
$f_{P}$ agree well, with the errors on $f_{P}^{(2)}$ being
smaller than $f_{P}^{(1)}$ in most cases. 

\begin{table}[t]
\centering
\caption{Pseudoscalar decay constants as computed  from the
  double-cosh fits to the  LL-LL  and GG-GG 
  axial vector correlation functions
  for all three ensembles.}
\label{tab:fpi_GL}
\begin{tabular*}{\columnwidth}{@{\extracolsep{\fill}}cccc}
\hline
\hline
$m_l^{\rm sea}$ & $m_{1}$  & $m_2$ & $f^{(1)}_{P}$ \\
\hline
0.01 &  0.01 & 0.01   &  0.0893(13)  \\
0.01 &  0.01 & 0.04   &  0.1007(12)  \\
0.01 &  0.04 & 0.04   &  0.1116(10)  \\
\hline
0.02 &  0.02 & 0.02   &  0.1003(8) \\   
0.02 &  0.02 & 0.04   &  0.1070(6) \\
0.02 &  0.04 & 0.04   &  0.1133(7) \\     
\hline
0.03 &  0.03 & 0.03   &  0.1094(13) \\
0.03 &  0.03 & 0.04   &  0.1111(12) \\
0.04 &  0.04 & 0.04   &  0.1143(12) \\      
\hline
\hline
\end{tabular*}
\end{table}

\begin{table}[t]
\centering
\caption{Pseudoscalar decay constants as computed  from the WL-WL 
  axial vector correlation functions
  for all three ensembles.}
\label{tab:fpi_WL}
\begin{tabular*}{\columnwidth}{@{\extracolsep{\fill}}cccc}
\hline
\hline

$m_l^{\rm sea}$ & $m_{1}$  & $m_2$ & $f^{(2)}_{P}$ \\
\hline
0.01 & 0.01 & 0.01 &  0.0895(7) \\
0.01 & 0.02 & 0.02 &  0.0977(7) \\
0.01 & 0.03 & 0.03 &  0.1046(7) \\
0.01 & 0.04 & 0.04 &  0.1108(7) \\
\hline
0.02 & 0.01 & 0.01 &  0.0939(7) \\
0.02 & 0.02 & 0.02 & 0.1006(6) \\
0.02 & 0.03 & 0.03 &  0.1067(6) \\
0.02 & 0.04 & 0.04 &  0.1123(6) \\
\hline
0.03 & 0.01 & 0.01 &  0.0980(6) \\
0.03 & 0.02 & 0.02 &  0.1042(6) \\
0.03 & 0.03 & 0.03 &  0.1101(6) \\
0.03 & 0.04 & 0.04 &  0.1157(6) \\
\hline
\hline
\end{tabular*}
\end{table}
\section{Observables in the Chiral Limit}
\label{sec:chiral_limit}
Since we perform our simulations at quark masses that are heavier than the
physical up and down quark masses, extrapolations are needed to obtain
physical values for various quantities. Chiral
perturbation theory is the correct effective
theory to describe low-energy QCD and should be used to 
guide the extrapolations. However, the quark masses in our simulations are 
too heavy for next-to-leading-order (NLO) chiral perturbation theory to
be consistent 
with our results~\cite{Lin:2006cf}. Although independent NLO fits for
the pseudoscalar meson masses and decay constants are consistent with
our data, our attempt to fit both simultaneously fails badly. This is
to be expected since the pion masses at the light dynamical points in
our simulations  range from 400 MeV to
627 MeV.  
This mass scale is likely to be outside of the region where chiral
perturbation  theory is
applicable. As will be demonstrated in the 
following, our data exhibits linear behavior in the range of quark 
masses in our simulations. Thus we resort to simple linear
extrapolations to  obtain
physical results in the light quark mass limit. In many cases this
coincides with the predictions of the leading-order chiral
perturbation theory, but in fact the reason we are doing this is
based purely on phenomenological observations, not from the underlying
effective  theory. 

\subsection{Determination of the Lattice Scale}
\begin{figure}[t]
\centering
\includegraphics[width=\columnwidth]{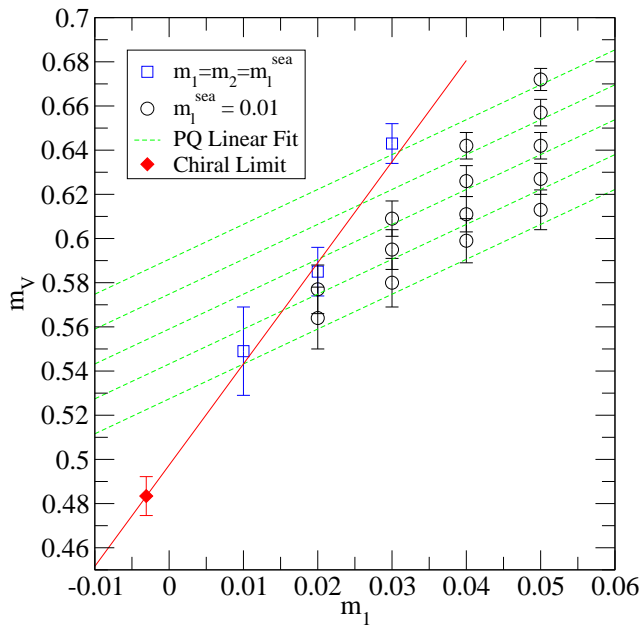}
\caption{Linear extrapolations of the vector mass using
Eq.(\ref{eq:rho_PQfit}). Here the circles are partially quenched data
points from the $m_l^{\rm sea} = 0.01$ ensemble, the squares are the
unitary points with $m_1 = m_2 = m_l^{\rm sea}$ and the diamond is the
value of $m_V$ in the chiral limit. }
\label{fig:rho_extra_0.01}
\end{figure}

We start out by setting the lattice scale  for these simulations from
the vector $\rho$ mass. Although at the lightest quark masses the
vector meson mass in our  simulations is slightly larger than the
2$m_\pi$ threshold,  the  requirement for non-zero relative momentum
for the two pions  prohibits the vector meson from decaying. We have
chosen to do a  partially quenched linear fit to the values 
of $m_{V}$ in Table~\ref{tab:summary_table}, restricted to $m_1(m_2)
\leq 0.03$,  with the following phenomenological form :
\begin{equation}
m_{V} = A\, (m_l^{\rm sea} + \mres) + B \, (m_1 + m_2 + 2\mres) + C.
\label{eq:rho_PQfit}
\end{equation}
The values of the fit parameters are shown in
Table~\ref{tab:chiral_limit}. As an illustration, we show the fit curves
through  the $m_l^{\rm sea} = 0.01$ data points {\footnote{The
$m_l^{\rm sea}$ = 0.02 and 0.03 results are also included in the fit, but
for clarity, we do not show them on the graph.}} and the unitary
points in Figure~\ref{fig:rho_extra_0.01}.  Setting $m_1 = m_2 =
m_l^{\rm sea} = -\mres$ gives the $\rho$ mass in the chiral limit, from 
which we determine  the lattice scale to be 
\begin{equation}
a^{-1} |_\rho= 1.61(3)\, {\rm GeV}. 
\end{equation}

We also determined the lattice scale from the static
quark potential using the Coulomb gauge
method~\cite{Bernard:2000gd}. The preliminary analysis of these
ensembles was reported in \cite{Li:2006gr}. More detailed analysis 
of this calculation is the subject of another
paper in progress and will not be addressed here. We simply quote the 
updated value of the lattice scale determined using $r_0 = 0.5$ fm, which is
\begin{equation}
a^{-1} |_{r_0}= 1.63(5)\, {\rm GeV}. 
\end{equation}

\begin{figure}[t]
\centering
\includegraphics[width=\columnwidth]{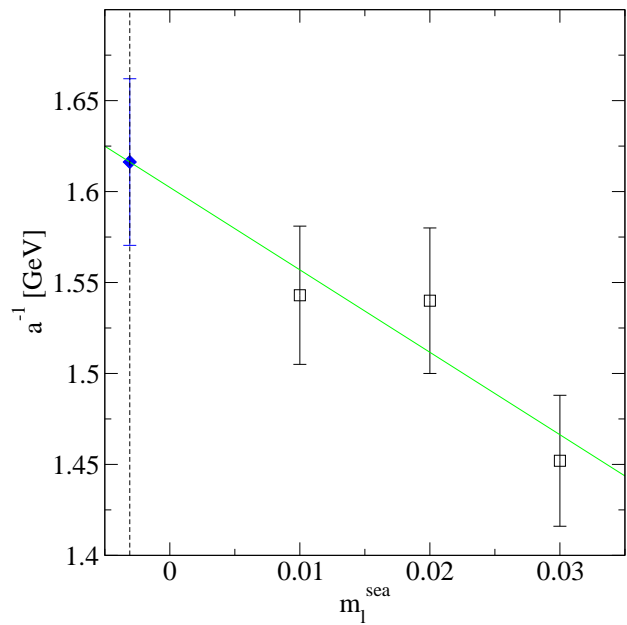}
\caption{Linear extrapolation of $a^{-1}$ versus $m_l^{\rm sea}$ from the
method of planes.}
\label{fig:inva_J_vs_msea}
\end{figure}

Our third choice for determining the lattice scale is the
``method of lattice planes'' \cite{Allton:1996yv}.
In this approach the vector meson mass is plotted against the
pseudoscalar meson mass squared and the interception of this data with
the physical point $(m_K^2,m_{K\ast})$ is found. Since the lattice
data straddles the kaon data, a chiral {\em interpolation} rather than
extrapolation is required in the valence quark sector. The results for
the inverse lattice spacing for each of the sea quark masses is shown
in Figure \ref{fig:inva_J_vs_msea}. Shown in this figure is a linear
extrapolation in the sea quark mass to the chiral limit, $m_l^{\rm sea} =
-\mres$. In this limit, we obtain

\begin{equation}
a^{-1}|_{\rm lattice\ plane} = 1.62(5)\ {\rm GeV}.
\end{equation}

All these methods give consistent results for the lattice scale.
However, each method has its own limitations. The physical $\rho$
meson is unstable and  has the probability of decaying into two pions
(though it is not  the case in our simulations). The static
quark potential suffers  a small (few percent) uncertainty
concerning the value of $r_0$. For the method of
lattice planes, the extrapolation in the sea sector is purely
phenomenological. Nevertheless, the three determinations of $a^{-1}$
agree well, so we take the average of these for our central value, and
take as an error an average of the statistical errors. This gives
\begin{equation}
a^{-1} = 1.62(4)\, {\rm GeV},
\end{equation}
which will be used  whenever a lattice scale is needed, and the errors 
will be propagated accordingly by quadrature. 

\subsection{Quark Masses and Decay Constants}
\begin{figure}[t]
\centering
\includegraphics[width=\columnwidth]{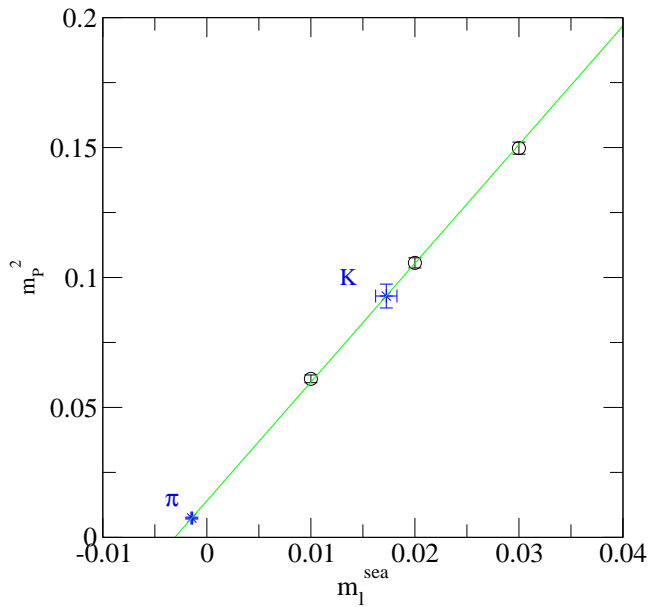}
\caption{Linear extrapolation of the pseudoscalar meson mass through
  the light dynamical points. The crosses are the physical $\pi$ and $K$
  points from which we obtain the values of $\bar{m}$ and $m_s$.  }
\label{fig:pion_extrap}
\end{figure}

A precise determination of the physical quark masses demands
well-controlled chiral extrapolations in the light quark
limit. As already mentioned, the relatively heavy quark masses in our
simulations prevent us from using the
next-to-leading order chiral perturbation
theory to do the extrapolations. Again we restrict ourselves to
linear extrapolation. We chose to fit only to the dynamical points
where $m_{\rm val} =
m_l^{\rm sea}$, using the following formula:
\begin{equation}
m_{P}^2 = 2 B (m_l^{\rm sea} + \mres).
\label{eq:pion_linear}
\end{equation}
The value of $B$ is given in Table~\ref{tab:chiral_limit}. We then
used the physical pion  and kaon masses as inputs to determine
the average light quark mass and the strange quark mass, as shown in
Figure~\ref{fig:pion_extrap}. The bare quark masses in lattice units are
listed in the following:
\begin{eqnarray}
\bar{m} & = & 0.00162(8) \\
m_s & = & 0.0390(21).  
\end{eqnarray}
The above results have combined the bare input quark masses and the residual
mass, and the quoted errors come from the errors on the
pseudoscalar meson masses, the residual mass, and the lattice scale. 
Systematic uncertainties
associated with the finite lattice spacing, chiral extrapolation
and \emph{e-m} splitting, etc.
have not been investigated yet. We will employ the
non-perturbative  renormalization (NPR)
technique~\cite{Blum:2001sr} to obtain the renormalized quark
masses. The NPR analysis is still in progress.

\begin{figure}[t]
\centering
\includegraphics[width=\columnwidth]{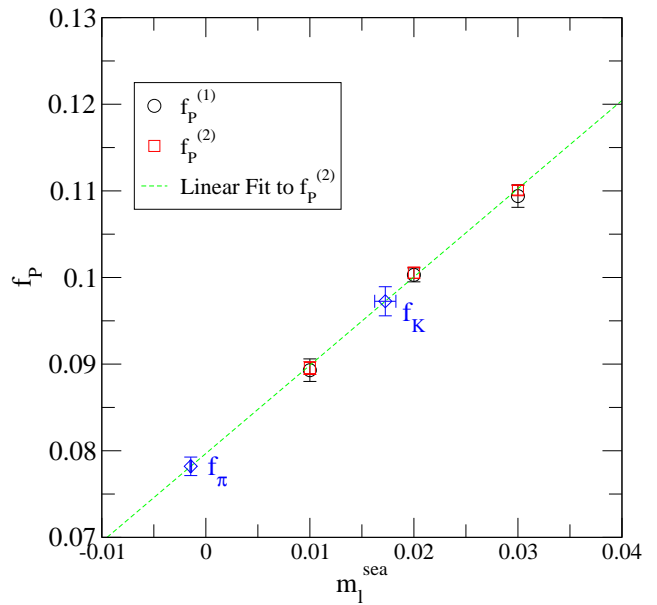}
\caption{Linear extrapolation of the pseudoscalar meson decay constant through
  the light dynamical points. The diamonds are the values of $f_\pi$
  and $f_K$ obtained from $f_{P}^{(2)}$.  }
\label{fig:fpi_extrap}
\end{figure}

To obtain the physical results of $f_\pi$ and $f_K$, we fit the results of the
pseudoscalar decay constants in Tables~\ref{tab:fpi_WL} and
\ref{tab:fpi_GL} to the following linear form separately:
\begin{equation}
f_{P} = f_0 + L (\, m_l^{\rm sea} + \mres).
\end{equation}
We can see from Figure~\ref{fig:fpi_extrap} that the data points are
quite linear with our current statistics. The fitting parameters are
also tabulated  in
Table~\ref{tab:chiral_limit}. Using the quark masses and lattice scale
determined in the  above,  we
found that the physical values of $f_\pi$, $f_K$ and $f_K/f_\pi$  to
be 126.9(45) MeV,  157.3(51) MeV and 1.240(31) respectively using
$f_{P}^{(1)}$. Similar analysis using $f_{P}^{(2)}$ gives
$f_\pi = 126.7(36)$ MeV, $f_K = 157.6(45)$ MeV and $f_K/f_\pi = 1.244(20)$, 
which are consistent with the results from $f_{P}^{(1)}$.  
We take the averages of these two methods as our best estimates for
these quantities, which are summarized in the following:
\begin{eqnarray}
f_\pi & = & 127(4)\,\, {\rm MeV}, \\
f_K & = & 157(5)\,\, {\rm MeV}, \\
f_K/f_\pi &=& 1.24(2).
\end{eqnarray}
Our convention for the pseudoscalar decay
constant is such that the experimental value of $f_\pi$ is about 131
MeV,  $f_K$ about 160 MeV and $f_K/f_\pi$ about 1.22. Thus our results
are in good agreement with the experiment. 

\begin{table}[b]
\centering
\caption{Linear fit parameters for various quantities.}
\begin{tabular*}{\columnwidth}{@{\extracolsep{\fill}}cccc}
 
\hline
\hline
$m_{V}$ &  $A=1.42(23)$ & $B=1.58(16)$  &  $C=0.483(9)$ \\
\hline
$m_{P}^2$   & & $B=2.285(24)$  & \\
\hline
$f_{P}^{(1)}$ & $L = 1.01(9)$ &  $f_0 = 0.0767(20)$  & \\
$f_{P}^{(2)}$ & $L = 1.02(5)$  &  $f_0 = 0.0765(11)$ & \\
\hline
\hline
\end{tabular*}
\label{tab:chiral_limit}
\end{table}

We stress that these physical results  are obtained in the context of
simple linear extrapolations. Systematic uncertainties may arise from
the presence of chiral logs. More sophisticated examination of the
chiral behavior of various observables requires simulations at
lighter quark masses, which will be further addressed in the near
future. Nevertheless, such agreement is encouraging, and demonstrates
that simulations with 2+1 flavors of domain wall fermions are promising.

\section{Comparison of RHMC I and  RHMC II}
\label{sec:quotient_RHMC}
As discussed in Section~\ref{sec:ensemble_properties}, there seem to be
some long-range autocorrelations in these ensembles generated with 2+1 
flavors of domain wall fermions. However, early simulations with two
flavors of dynamical domain wall fermions did not have this
problem~\cite{Aoki:2004ht}. The biggest difference between those
two-flavor simulations  and this work lies in the way the following ratio
is evaluated, 
\begin{equation}
\frac{
 \det \left[ D_{\rm DWF}^\dagger(M_5, m_s) D_{\rm DWF}(M_5, m_s)
   \right] }{
 \det \left[ D_{\rm DWF}^\dagger(M_5, 1) D_{\rm DWF}(M_5, 1)
    \right]
 }. 
 \label{eq:nf2_det}
\end{equation}
In the two-flavor case, this ratio was evaluated using one
pseudo-fermion field in the HMC evolution~\cite{Aoki:2004ht}, which was found to reduce
the stochastic noise substantially, and a larger step size could be
used. At the beginning  of these
simulations this method had not yet been implemented in the RHMC
algorithm, and two separate pseudo-fermion fields were needed to
evaluate Eq.~(\ref{eq:nf2_det}). Upon completion of the work reported in
the previous sections, the new code was ready, and we were motivated to
perform an extension of the $m_l^{\rm sea} = 0.03$ ensemble using this
new code to investigate any effects it might have on the ensemble
properties.  This variant of the RHMC algorithm, RHMC II, has
been described in detail in Section~\ref{sec:simulation_details}. From
Table~\ref{tab:run-details} we can see that the coarsest step size,
$\delta\tau$,  of RHMC II is almost twice as large as RHMC I, and
fewer steps of intermediate and finest integration sizes are
needed. We even obtained higher acceptance with these
parameters. There is clearly at least a factor of two speedup by
switching from RHMC I to RHMC II.

\begin{figure}[t]
\centering
\includegraphics[width=\columnwidth]{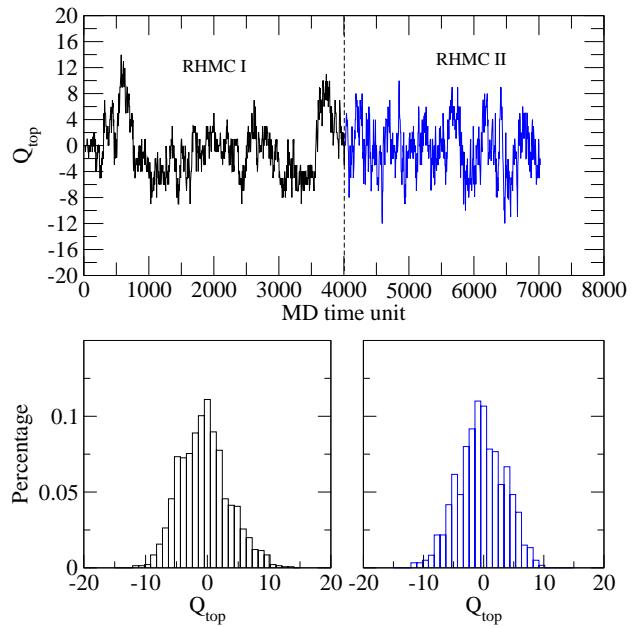}
\caption{The comparison of the topological charge evolution and
  corresponding histogram  measured from the two portions of the
  $m_l^{\rm sea} = 0.03$  ensemble, in which different algorithms
  (RHMC I and RHMC II) were  used to evaluate
  Eq.(\ref{eq:nf3-det}). The dashed line indicates  where the algorithm was changed. }
\label{fig:new_topology}
\end{figure}

\begin{figure}[htb]
\centering
\includegraphics[width=\columnwidth]{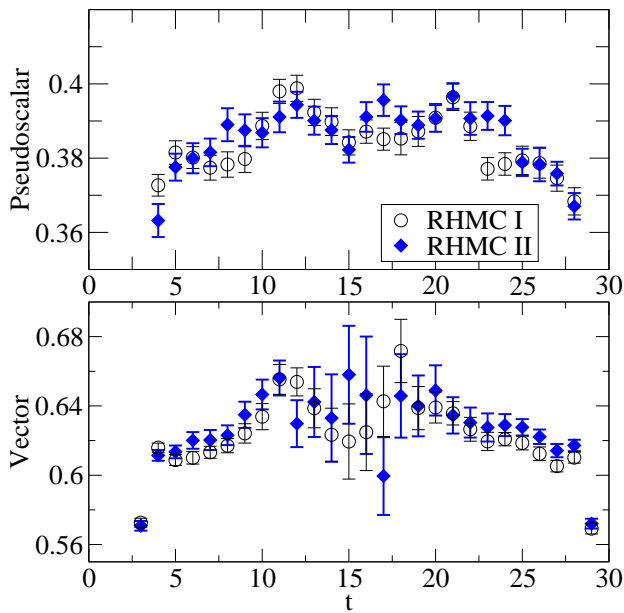}
\caption{The comparison of the effective masses  for the pseudoscalar
  and vector WL-WL correlation  functions at the unitary point from
  the two portions of the  $m_l^{\rm sea} = 0.03$ ensemble  (RHMC I and RHMC II). }
\label{fig:new_effm}
\end{figure}

Figure~\ref{fig:new_topology} shows the comparison of the evolution of
the topological charge on the lattices generated using RHMC I and RHMC
II. The tunneling of topological charge in RHMC II appears to be much
better than RHMC I. The long-range autocorrelations present in RHMC I
do not show up in RHMC II.   This can be partly attributed to the use
of a larger step size  in the molecular dynamics evolution.
Intuitively, a larger step size allows more efficient movement
through, and the evolution is less  likely to be trapped in a
small region of, phase space. This consequently helps to move the
topological charge from one sector to another.  However, the histograms of the
topological charge from these two segments of the ensemble are quite
similar, indicating
that RHMC I, while showing longer range autocorrelations, does
give the same results as the improved version of the algorithm, RHMC II. We
have also seen large fluctuations of the effective masses in
Section~\ref{sec:observables}. As a comparison, we show the effective
masses for the pseudoscalar and vector WL-WL correlation functions,
measured every 20 trajectories with  $t_{\rm src} = 0$  at the unitary
point in Figure~\ref{fig:new_effm}. While the vector effective mass
remains noisy for RHMC II, there is noticeable improvement in the
pseudoscalar effective mass, where a better plateau can be identified
from $t = 8$ to $t = 24$. 

We believe that RHMC II not only speeds up the gauge
field generation by more than a factor of 2, but also helps to move through
phase space more quickly and improve the quality of the
ensembles we generate. The interplay of these two factors makes it a more
favorable algorithm to use than RHMC I.
\section{Conclusions}
\label{sec:conclusions}
We have presented the results of light meson physics from lattice
simulations with 2+1 flavors of  dynamical domain wall fermions at a
lattice spacing of about  0.12 fm in a (2 fm)$^3$ volume.  We used the
improved RHMC algorithm  to generate the gauge configurations, which
turned out to be quite  efficient in reducing the computing
cost. Although there are some  long-range autocorrelations in these
ensembles, we have been  able to obtain physical results with better
than 5\% statistical  accuracy.  The residual chiral symmetry breaking
in these calculations  is small compared to the earlier calculation
with $L_s = 8$~\cite{Antonio:2006px},  which gives us better control
over chiral symmetry breaking effects.  We have shown results for the
bare physical quark masses and decay  constants, the latter in good
agreement with their experimental values.  We
have also shown that a newly improved variant of the RHMC algorithm,
RHMC II, not only speeds up the gauge  field generation by more than a
factor of 2, but also moves through phase space more efficiently. 

There are of course some systematic uncertainties that we have yet to
control. One is the failure of even our lightest quark mass
data to permit chiral extrapolations which follow the 
form predicted by chiral perturbation theory. We did not attempt to
perform simulations at even lighter quark masses on these lattices, because
the volume for such simulations may be too small, which could
result in sizable finite volume effects. To pursue this issue
further, the RBC and UKQCD collaborations are performing 2+1 flavor
simulations on a $24^3\times64$ volume, or about
(3 fm)$^3$ in physical units, at about the same lattice spacing. This
larger  volume allows
calculations with lighter quark masses, from which we hope to
investigate the light quark limit in the context
of chiral perturbation theory. Comparing the physical results in two
different volumes will also enable us to quantitatively study finite volume
effects. This will be especially important for baryon
physics. Secondly, more  than one lattice spacing
is needed to investigate the scaling behavior for domain wall fermions
and the Iwasaki gauge action, and the continuum limit should be taken
to remove lattice artifacts. We will address this question in
upcoming simulations in a (3 fm)$^3$ volume at a weaker coupling. 

With good control over chiral symmetry and an exact and
efficient algorithm, we are vigorously pursuing 2+1 flavor domain wall
QCD. The theoretical promise of this approach is now producing
numerical results for the physically interesting 2+1 flavor case. 

\section*{Acknowledgments}
We thank Dong Chen, Calin Cristian, Zhihua Dong, Alan Gara, Andrew
Jackson, Changhoan Kim, Ludmila Levkova, Xiaodong Liao, Guofeng
Liu, Konstantin Petrov and Tilo Wettig for developing with us the QCDOC
machine and its software. This development and the resulting computer
equipment used in this calculation were funded by the U.S.\ DOE grant
DE-FG02-92ER40699, PPARC JIF grant PPA/J/S/1998/00756 and by RIKEN. This work
was supported by DOE grants DE-FG02-92ER40699 and DE-AC02-98CH10886
and PPARC grants PPA/G/O/2002/00465, PP/D000238/1, PP/C504386/1,
PPA/G/S/2002/00467 and PP/D000211/1. 
AH is supported by the U.K. Royal Society. 
We thank BNL, EPCC, RIKEN, and the U.S.\ DOE for
supporting the computing facilities essential for the completion of
this work.

\bibliography{paper}

\appendix
\pagebreak
\section[floatfix]{Tables of Meson Masses}
\label{sec:appendix_tables}
\begin{table}[htbp]
\centering
\caption{Pseudoscalar meson masses from GL-GL correlation functions
  as determined from the pseudoscalar channel.}
\label{tab:pseudo_GL}
\begin{tabular*}{\columnwidth}{@{\extracolsep{\fill}}cccccc}
\hline
\hline
$m_l^{\rm sea}$ & $m_1$ & $m_2$ & $t_{\rm min}-t_{\rm max}$ & $\chi^2$/dof & $m_{P}$ \\
\hline
0.01 & 0.01 & 0.01 & 8-16 &  0.6(6)  & 0.2460(10) \\
0.01 & 0.01 & 0.04 & 8-16 &  0.8(6)  & 0.3545(8) \\
0.01 & 0.04 & 0.04 & 8-16 &  1.1(7)  & 0.4382(7) \\
\hline
0.02 & 0.02 & 0.02 & 8-16 &  0.9(7)  & 0.3247(8) \\
0.02 & 0.02 & 0.04 & 8-16 &  0.9(7)  & 0.3868(8) \\
0.02 & 0.04 & 0.04 & 8-16 &  0.9(7)  & 0.4410(7) \\
\hline
0.03 & 0.03 & 0.03 & 8-16 &  3.8(15)  & 0.3907(15) \\
0.03 & 0.03 & 0.04 & 8-16 &  3.6(14)  & 0.4184(14) \\
0.03 & 0.04 & 0.04 & 8-16 &  3.5(14)  & 0.4446(13) \\
\hline
\hline
\end{tabular*}
\end{table}

\begin{table}[htbp]
\centering
\caption{Pseudoscalar meson masses from GL-GL correlation functions
  as determined from the axial vector channel.}
\label{tab:axialt_GL}
\begin{tabular*}{\columnwidth}{@{\extracolsep{\fill}}cccccc}
\hline
\hline
$m_l^{\rm sea}$ & $m_1$ & $m_2$ & $t_{\rm min}-t_{\rm max}$ & $\chi^2$/dof & $m_{P}$ \\
\hline
0.01 & 0.01 & 0.01 & 8-16 & 2.4(10) & 0.2467(15) \\
0.01 & 0.01 & 0.04 & 8-16 & 1.9(8) & 0.3547(14) \\
0.01 & 0.04 & 0.04 & 8-16 & 2.0(9) & 0.4379(10) \\
\hline
0.02 & 0.02 & 0.02 & 9-16 & 2.1(10) & 0.3254(12)  \\
0.02 & 0.02 & 0.04 & 9-16 & 1.8(9) & 0.3877(10)  \\
0.02 & 0.04 & 0.04 & 9-16 & 1.7(9) & 0.4413(10)  \\
\hline
0.03 & 0.03 & 0.03 & 7-16 & 3.2(11) & 0.3881(16)  \\
0.03 & 0.03 & 0.04 & 7-16 & 2.8(10) & 0.4170(16)  \\
0.03 & 0.04 & 0.04 & 7-16 & 1.7(9) & 0.4425(13)  \\
\hline
\hline
\end{tabular*}
\end{table}

\begin{table}[htbp]
\centering
\caption{Vector meson masses from GL-GL correlation functions
  for all three ensembles.}
\label{tab:vector_GL}
\begin{tabular*}{\columnwidth}{@{\extracolsep{\fill}}cccccc}
\hline
\hline
$m_l^{\rm sea}$ & $m_{1}$  & $m_2$ & $t_{\rm min}-t_{\rm max}$ & $\chi^2/dof$ & $m_{V}$ \\
\hline
0.01 & 0.01 & 0.01 & 10-16 &  0.7(7)  & 0.558(12) \\
0.01 & 0.01 & 0.04 & 10-16 &  0.6(6)  & 0.6016(45) \\
0.01 & 0.04 & 0.04 & 10-16 &  0.7(7)  & 0.6443(23) \\
\hline
0.02 & 0.02 & 0.02 & 10-16 &  1.4(10)  & 0.5914(69) \\
0.02 & 0.02 & 0.04 & 10-16 &  0.7(8)  & 0.6217(47) \\
0.02 & 0.04 & 0.04 & 10-16 &  0.4(6)  & 0.6524(35) \\
\hline
0.03 & 0.03 & 0.03 & 10-16 &  0.7(8)  & 0.6519(75) \\
0.03 & 0.03 & 0.04 & 10-16 &  0.7(7)  & 0.6643(63) \\
0.03 & 0.04 & 0.04 & 10-16 &  0.6(7)  & 0.6771(54) \\
\hline
\hline
\end{tabular*}
\end{table}

\begin{table}[htbp]
\centering
\caption{Pseudoscalar meson masses  as determined from the WL-WL
 pseudoscalar correlation functions.}
\label{tab:pseudo_WL}
\begin{tabular*}{\columnwidth}{@{\extracolsep{\fill}}cccccc}
\hline
\hline
$m_l^{\rm sea}$ & $m_1$ & $m_2$ & $t_{\rm min}-t_{\rm max}$ & $\chi^2$/dof & $m_{P}$ \\
\hline
0.01 & 0.01 & 0.01 & 9-16 & 1.1(9) & 0.2474(16) \\
0.01 & 0.02 & 0.02 & 9-16 & 1.3(9) & 0.3223(14) \\
0.01 & 0.03 & 0.03 & 9-16 & 1.4(10) & 0.3836(12) \\
0.01 & 0.04 & 0.04 & 9-16 & 1.6(10) & 0.4373(11) \\
\hline
0.02 & 0.01 & 0.01 & 9-16 & 1.1(8) & 0.2458(15) \\
0.02 & 0.02 & 0.02 & 9-16 & 1.3(9) & 0.3212(12) \\
0.02 & 0.03 & 0.03 & 9-16 & 1.6(10) & 0.3832(11) \\
0.02 & 0.04 & 0.04 & 9-16 & 1.9(11) & 0.4378(10) \\
\hline
0.03 & 0.01 & 0.01 & 9-16 & 3.5(15) & 0.2551(19) \\
0.03 & 0.02 & 0.02 & 9-16 & 3.6(15) & 0.3282(16)\\
0.03 & 0.03 & 0.03 & 9-16 & 3.5(15) & 0.3891(15) \\
0.03 & 0.04 & 0.04 & 9-16 & 3.4(15) & 0.4429(13) \\
\hline
\hline
\end{tabular*}
\end{table}

\begin{table}[htbp]
\centering
\caption{Pseudoscalar meson masses from WL-WL correlation functions
  as determined from the axial vector channel.}
\label{tab:axialt_WL}
\begin{tabular*}{\columnwidth}{@{\extracolsep{\fill}}cccccc}
\hline
\hline
$m_l^{\rm sea}$ & $m_{1}$  & $m_2$ & $t_{\rm min}-t_{\rm max}$ & $\chi^2$/dof& $m_{P}$ \\
\hline
0.01 & 0.01 & 0.01 & 8-16 &  0.8(7)  & 0.2466(16) \\
0.01 & 0.02 & 0.02 & 8-16 &  0.6(6)  & 0.3215(14) \\
0.01 & 0.03 & 0.03 & 8-16 &  0.8(7)  & 0.3828(13) \\
0.01 & 0.04 & 0.04 & 8-16 &  1.2(8)  & 0.4365(12) \\
\hline
0.02 & 0.01 & 0.01 & 8-16 &  1.4(9)  & 0.2479(13) \\
0.02 & 0.02 & 0.02 & 8-16 &  1.6(10)  & 0.3226(11) \\
0.02 & 0.03 & 0.03 & 8-16 &  1.7(10)  & 0.3842(10) \\
0.02 & 0.04 & 0.04 & 8-16 &  1.9(10)  & 0.4385(9) \\
\hline
0.03 & 0.01 & 0.01 & 8-16 &  2.2(11)  & 0.2526(14) \\
0.03 & 0.02 & 0.02 & 8-16 &  3.0(13)  & 0.3268(14) \\
0.03 & 0.03 & 0.03 & 8-16 &  2.9(13)  & 0.3884(13) \\
0.03 & 0.04 & 0.04 & 8-16 &  2.8(13)  & 0.4427(11) \\
\hline
\hline
\end{tabular*}
\end{table}

\begin{table}[htbp]
\centering
\caption{Vector meson masses from WL-WL correlation functions
  for all three ensembles.}
\label{tab:vector_WL}
\begin{tabular*}{\columnwidth}{@{\extracolsep{\fill}}cccccc}
\hline
\hline
$m_l^{\rm sea}$ & $m_{1}$  & $m_2$ & $t_{\rm min}-t_{\rm max}$ & $\chi^2/dof$ & $m_{V}$ \\
\hline
0.01 & 0.01 & 0.01 & 9-16 &  1.2(9)  & 0.542(11) \\
0.01 & 0.02 & 0.02 & 9-16 &  1.9(11)  & 0.5742(58) \\
0.01 & 0.03 & 0.03 & 9-16 &  1.9(11)  & 0.6078(38) \\
0.01 & 0.04 & 0.04 & 9-16 &  1.6(10)  & 0.6402(29) \\
\hline
0.02 & 0.01 & 0.01 & 6-16 &  0.8(6)  & 0.5533(44) \\
0.02 & 0.02 & 0.02 & 6-16 &  0.8(6)  & 0.5845(30) \\
0.02 & 0.03 & 0.03 & 6-16 &  0.9(6)  & 0.6162(23) \\
0.02 & 0.04 & 0.04 & 6-16 &  1.1(7)  & 0.6475(19) \\
\hline
0.03 & 0.01 & 0.01 & 10-16 &  0.9(8)  & 0.621(16) \\
0.03 & 0.02 & 0.02 & 10-16 &  0.8(8)  & 0.6253(82) \\
0.03 & 0.03 & 0.03 & 10-16 &  1.0(9)  & 0.6464(55) \\
0.03 & 0.04 & 0.04 & 10-16 &  1.4(11)  & 0.6718(42) \\
\hline
\hline
\end{tabular*}
\end{table}

\begin{table}[htbp]
\centering
\caption{Pseudoscalar meson masses from WL-WL correlation functions
  with P+A boundary conditions as determined from the pseudoscalar channel.}
\label{tab:pseudo_PA}
\begin{tabular*}{\columnwidth}{@{\extracolsep{\fill}}ccccc}
\hline
\hline
$m_l^{\rm sea}$ & $m_{1}$  & $m_2 $& $t_{\rm min}-t_{\rm max}$ & $m_{P}$ \\
\hline
0.01 & 0.01 & 0.01 & 10-22 & 0.2509(22) \\
0.01 & 0.02 & 0.01 & 10-22 & 0.2913(21) \\
0.01 & 0.02 & 0.02 & 10-22 & 0.3262(19) \\
0.01 & 0.03 & 0.01 & 10-22 & 0.3270(20) \\
0.01 & 0.03 & 0.02 & 10-22 & 0.3583(19) \\
0.01 & 0.03 & 0.03 & 10-22 & 0.3877(18) \\
0.01 & 0.04 & 0.01 & 10-22 & 0.3594(20) \\
0.01 & 0.04 & 0.02 & 10-22 & 0.3881(18) \\
0.01 & 0.04 & 0.03 & 10-22 & 0.4155(18) \\
0.01 & 0.04 & 0.04 & 10-22 & 0.4416(17) \\
0.01 & 0.05 & 0.01 & 10-22 & 0.3894(20) \\
0.01 & 0.05 & 0.02 & 10-22 & 0.4161(18) \\
0.01 & 0.05 & 0.03 & 10-22 & 0.4418(18) \\
0.01 & 0.05 & 0.04 & 10-22 & 0.4666(17) \\
0.01 & 0.05 & 0.05 & 10-22 & 0.4906(17) \\
\hline
0.02 & 0.01 & 0.01 & 10-22 & 0.2527(20) \\
0.02 & 0.02 & 0.01 & 10-22 & 0.2910(19) \\
0.02 & 0.02 & 0.02 & 10-22 & 0.3251(18) \\
0.02 & 0.03 & 0.01 & 10-22 & 0.3253(19) \\
0.02 & 0.03 & 0.02 & 10-22 & 0.3564(17) \\
0.02 & 0.03 & 0.03 & 10-22 & 0.3855(17) \\
0.02 & 0.04 & 0.01 & 10-22 & 0.3567(19) \\
0.02 & 0.04 & 0.02 & 10-22 & 0.3856(17) \\
0.02 & 0.04 & 0.03 & 10-22 & 0.4129(16) \\
0.02 & 0.04 & 0.04 & 10-22 & 0.4390(15) \\
0.02 & 0.05 & 0.01 & 10-22 & 0.3859(19) \\
0.02 & 0.05 & 0.02 & 10-22 & 0.4131(17) \\
0.02 & 0.05 & 0.03 & 10-22 & 0.4390(16) \\
0.02 & 0.05 & 0.04 & 10-22 & 0.4639(15) \\
0.02 & 0.05 & 0.05 & 10-22 & 0.4878(15) \\
\hline
0.03 & 0.01 & 0.01 & 10-22 & 0.2496(21) \\
0.03 & 0.02 & 0.01 & 10-22 & 0.2885(19) \\
0.03 & 0.02 & 0.02 & 10-22 & 0.3227(19) \\
0.03 & 0.03 & 0.01 & 10-22 & 0.3235(19) \\
0.03 & 0.03 & 0.02 & 10-22 & 0.3544(18) \\
0.03 & 0.03 & 0.03 & 10-22 & 0.3837(18) \\
0.03 & 0.04 & 0.01 & 10-22 & 0.3556(19) \\
0.03 & 0.04 & 0.02 & 10-22 & 0.3841(18) \\
0.03 & 0.04 & 0.03 & 10-22 & 0.4116(17) \\
0.03 & 0.04 & 0.04 & 10-22 & 0.4379(17) \\
0.03 & 0.05 & 0.01 & 10-22 & 0.3854(19) \\
0.03 & 0.05 & 0.02 & 10-22 & 0.4120(18) \\
0.03 & 0.05 & 0.03 & 10-22 & 0.4380(17) \\
0.03 & 0.05 & 0.04 & 10-22 & 0.4631(17) \\
0.03 & 0.05 & 0.05 & 10-22 & 0.4873(16) \\
\hline
\hline
\end{tabular*}
\end{table}

\begin{table}[htbp]
\centering
\caption{Pseudoscalar meson masses from WL-WL correlation functions
  with P+A boundary conditions as determined from the axial vector
  channel.}
\label{tab:axialt_PA}
\begin{tabular*}{\columnwidth}{@{\extracolsep{\fill}}ccccc}
\hline
\hline
$m_l^{\rm sea}$ & $m_{1}$  & $m_2$ & $t_{\rm min}-t_{\rm max}$  & $m_{P}$ \\
\hline
0.01 & 0.01 & 0.01 & 10-22 & 0.2473(23) \\
0.01 & 0.02 & 0.01 & 10-22 & 0.2878(21) \\
0.01 & 0.02 & 0.02 & 10-22 & 0.3231(19) \\
0.01 & 0.03 & 0.01 & 10-22 & 0.3236(20) \\
0.01 & 0.03 & 0.02 & 10-22 & 0.3552(18) \\
0.01 & 0.03 & 0.03 & 10-22 & 0.3847(17) \\
0.01 & 0.04 & 0.01 & 10-22 & 0.3560(20) \\
0.01 & 0.04 & 0.02 & 10-22 & 0.3850(17) \\
0.01 & 0.04 & 0.03 & 10-22 & 0.4124(16) \\
0.01 & 0.04 & 0.04 & 10-22 & 0.4386(15) \\
0.01 & 0.05 & 0.01 & 10-22 & 0.3860(19) \\
0.01 & 0.05 & 0.02 & 10-22 & 0.4130(17) \\
0.01 & 0.05 & 0.03 & 10-22 & 0.4388(16) \\
0.01 & 0.05 & 0.04 & 10-22 & 0.4637(15) \\
0.01 & 0.05 & 0.05 & 10-22 & 0.4877(15) \\
\hline
0.02 & 0.01 & 0.01 & 10-22 & 0.2535(22) \\
0.02 & 0.02 & 0.01 & 10-22 & 0.2938(20) \\
0.02 & 0.02 & 0.02 & 10-22 & 0.3287(19) \\
0.02 & 0.03 & 0.01 & 10-22 & 0.3292(20) \\
0.02 & 0.03 & 0.02 & 10-22 & 0.3605(18) \\
0.02 & 0.03 & 0.03 & 10-22 & 0.3897(17) \\
0.02 & 0.04 & 0.01 & 10-22 & 0.3612(19) \\
0.02 & 0.04 & 0.02 & 10-22 & 0.3899(17) \\
0.02 & 0.04 & 0.03 & 10-22 & 0.4171(16) \\
0.02 & 0.04 & 0.04 & 10-22 & 0.4430(16) \\
0.02 & 0.05 & 0.01 & 10-22 & 0.3907(19) \\
0.02 & 0.05 & 0.02 & 10-22 & 0.4175(17) \\
0.02 & 0.05 & 0.03 & 10-22 & 0.4431(16) \\
0.02 & 0.05 & 0.04 & 10-22 & 0.4677(15) \\
0.02 & 0.05 & 0.05 & 10-22 & 0.4914(15) \\
\hline
0.03 & 0.01 & 0.01 & 10-22 & 0.2486(23) \\
0.03 & 0.02 & 0.01 & 10-22 & 0.2886(22) \\
0.03 & 0.02 & 0.02 & 10-22 & 0.3235(21) \\
0.03 & 0.03 & 0.01 & 10-22 & 0.3242(22) \\
0.03 & 0.03 & 0.02 & 10-22 & 0.3555(20) \\
0.03 & 0.03 & 0.03 & 10-22 & 0.3849(19) \\
0.03 & 0.04 & 0.01 & 10-22 & 0.3565(22) \\
0.03 & 0.04 & 0.02 & 10-22 & 0.3852(20) \\
0.03 & 0.04 & 0.03 & 10-22 & 0.4126(19) \\
0.03 & 0.04 & 0.04 & 10-22 & 0.4388(18) \\
0.03 & 0.05 & 0.01 & 10-22 & 0.3863(22) \\
0.03 & 0.05 & 0.02 & 10-22 & 0.4131(20) \\
0.03 & 0.05 & 0.03 & 10-22 & 0.4389(19) \\
0.03 & 0.05 & 0.04 & 10-22 & 0.4637(18) \\
0.03 & 0.05 & 0.05 & 10-22 & 0.4877(17) \\
\hline
\hline
\end{tabular*}
\end{table}

\begin{table}[htbp]
\centering
\caption{Vector meson masses from WL-WL correlation functions
  with P+A boundary conditions for all three ensembles.}
\label{tab:vector_PA}
\begin{tabular*}{\columnwidth}{@{\extracolsep{\fill}}ccccc}
\hline
\hline
$m_l^{\rm sea}$ & $m_{1}$  & $m_2$ & $t_{\rm min}-t_{\rm max}$  & $m_{V}$ \\
\hline
0.01 & 0.01 & 0.01 & 6-14 & 0.546(13) \\
0.01 & 0.02 & 0.01 & 6-14 & 0.563(9) \\
0.01 & 0.02 & 0.02 & 6-14 & 0.579(7) \\
0.01 & 0.03 & 0.01 & 6-14 & 0.580(7) \\
0.01 & 0.03 & 0.02 & 6-14 & 0.595(6) \\
0.01 & 0.03 & 0.03 & 6-14 & 0.610(5) \\
0.01 & 0.04 & 0.01 & 6-14 & 0.597(6) \\
0.01 & 0.04 & 0.02 & 6-14 & 0.611(5) \\
0.01 & 0.04 & 0.03 & 6-14 & 0.626(5) \\
0.01 & 0.04 & 0.04 & 6-14 & 0.641(4) \\
0.01 & 0.05 & 0.01 & 6-14 & 0.613(6) \\
0.01 & 0.05 & 0.02 & 6-14 & 0.627(5) \\
0.01 & 0.05 & 0.03 & 6-14 & 0.642(4) \\
0.01 & 0.05 & 0.04 & 6-14 & 0.657(4) \\
0.01 & 0.05 & 0.05 & 6-14 & 0.672(3) \\
\hline
0.02 & 0.01 & 0.01 & 6-14 & 0.538(13) \\
0.02 & 0.02 & 0.01 & 6-14 & 0.560(10) \\
0.02 & 0.02 & 0.02 & 6-14 & 0.579(7) \\
0.02 & 0.03 & 0.01 & 6-14 & 0.579(8) \\
0.02 & 0.03 & 0.02 & 6-14 & 0.597(6) \\
0.02 & 0.03 & 0.03 & 6-14 & 0.614(5) \\
0.02 & 0.04 & 0.01 & 6-14 & 0.597(7) \\
0.02 & 0.04 & 0.02 & 6-14 & 0.614(5) \\
0.02 & 0.04 & 0.03 & 6-14 & 0.631(5) \\
0.02 & 0.04 & 0.04 & 6-14 & 0.647(4) \\
0.02 & 0.05 & 0.01 & 6-14 & 0.614(6) \\
0.02 & 0.05 & 0.02 & 6-14 & 0.631(5) \\
0.02 & 0.05 & 0.03 & 6-14 & 0.647(4) \\
0.02 & 0.05 & 0.04 & 6-14 & 0.663(4) \\
0.02 & 0.05 & 0.05 & 6-14 & 0.679(4) \\
\hline
0.03 & 0.01 & 0.01 & 6-14 & 0.577(15) \\
0.03 & 0.02 & 0.01 & 6-14 & 0.589(11) \\
0.03 & 0.02 & 0.02 & 6-14 & 0.601(8) \\
0.03 & 0.03 & 0.01 & 6-14 & 0.603(9) \\
0.03 & 0.03 & 0.02 & 6-14 & 0.616(7) \\
0.03 & 0.03 & 0.03 & 6-14 & 0.630(6) \\
0.03 & 0.04 & 0.01 & 6-14 & 0.618(7) \\
0.03 & 0.04 & 0.02 & 6-14 & 0.631(6) \\
0.03 & 0.04 & 0.03 & 6-14 & 0.645(5) \\
0.03 & 0.04 & 0.04 & 6-14 & 0.660(4) \\
0.03 & 0.05 & 0.01 & 6-14 & 0.633(6) \\
0.03 & 0.05 & 0.02 & 6-14 & 0.646(5) \\
0.03 & 0.05 & 0.03 & 6-14 & 0.660(5) \\
0.03 & 0.05 & 0.04 & 6-14 & 0.675(4) \\
0.03 & 0.05 & 0.05 & 6-14 & 0.690(4) \\
\hline
\hline
\end{tabular*}
\end{table}
\clearpage
\end{document}